\documentclass[11pt,a4paper]{article}
\usepackage{jheppub}

\usepackage{amsfonts,amssymb}
\usepackage{amsmath}
\usepackage{epsfig, graphicx}

\newcommand{\be}{\begin{equation}}
\newcommand{\ee}{\end{equation}}
\newcommand{\bea}{\begin{eqnarray}}
\newcommand{\eea}{\end{eqnarray}}
\newcommand{\ba}{\begin{eqnarray}}
\newcommand{\ea}{\end{eqnarray}}
\newcommand{\nn}{\nonumber }

\newcommand{\beq}{\begin{equation}}
\newcommand{\eeq}{\end{equation}}
\newcommand{\beqa}{\begin{eqnarray}}
\newcommand{\eeqa}{\end{eqnarray}}
\newcommand{\beqar}{\begin{eqnarray*}}
\newcommand{\eeqar}{\end{eqnarray*}}

\def\t{\tilde}

\def\d{\partial}

\title{Effective actions for relativistic fluids from holography}

\author[a]{Jan de Boer,}
\author[b,*]{Michal P. Heller
\note[*]{On leave from: \emph{National Centre for Nuclear Research, Ho{\.z}a 69, 00-681 Warsaw, Poland}.}}
\author[a]{and Natalia Pinzani-Fokeeva}

 \affiliation[a]{Institute of Physics, Universiteit van Amsterdam,\\
 Science Park 904, 1090 GL Amsterdam, The Netherlands}

 \affiliation[b]{Perimeter Institute for Theoretical Physics,\\
 31 Caroline Street North, Waterloo, Ontario N2L 2Y5, Canada}

\emailAdd{j.deboer@uva.nl}
\emailAdd{mheller@pitp.ca}
\emailAdd{n.pinzanifokeeva@uva.nl}

\abstract{
Motivated by recent progress in developing action formulations of relativistic hydrodynamics, we use holography to derive the low energy dissipationless effective action for strongly coupled conformal fluids. Our analysis is based on the study of novel double Dirichlet problems for the gravitational field, in which the boundary conditions are set on two codimension one timelike hypersurfaces (branes). We provide a geometric interpretation of the Goldstone bosons appearing in such constructions in terms of a family of spatial geodesics extending between the ultraviolet and the infrared brane. Furthermore, we discuss supplementing double Dirichlet problems with information about the near-horizon geometry. We show that upon coupling to a membrane paradigm boundary condition, our approach reproduces correctly the complex dispersion relation for both sound and shear waves. We also demonstrate that upon a Wick rotation, our formulation reproduces the equilibrium partition function formalism, provided the near-horizon geometry is properly \mbox{accounted for}. Finally, we define the conserved hydrodynamic entropy current as the Noether current associated with a particular transformation of the Goldstone bosons.
}

\begin{document}  

\maketitle

\section{Introduction and summary}

The fact that hydrodynamics and gravity are closely connected to each other has been appreciated already for quite some time. Such a relation emerges naturally in the context of the AdS/CFT correspondence, as explored in the pioneering papers \cite{Policastro:2001yc,Policastro:2002se,Policastro:2002tn} and subsequently elucidated with the discovery of the fluid/gravity duality  \cite{Bhattacharyya:2008jc}, see also \cite{Rangamani:2009xk} for a review. One of the purposes of the current work is to provide an alternative derivation of hydrodynamics from gravity, focusing on the information encoded in the gravitational action.

The conventional description of relativistic hydrodynamics is given in terms of a conserved energy-momentum tensor and an entropy current, see, e.g., \cite{Landau6}. Recently, there has been a lot of progress in understanding this standard formulation from an action principle point of view. 
One of the main aims of these approaches is to  derive the hydrodynamic constitutive relations for the energy-momentum tensor without directly imposing the entropy production constraint.

One example is the partition function formalism for a fluid in thermodynamic equilibrium, coupled to an arbitrary background metric 
and possibly gauge fields \cite{Banerjee:2012iz,Jensen:2012jh}.
Since this approach is limited to thermal equilibrium, only dissipationless hydrostatic features are relevant. Quite impressively, it has been shown that the number of hydrostatic transport coefficients is in perfect agreement with the conventional approach after imposing the existence of an equilibrium configuration, see also \cite{Bhattacharyya:2014bha}.
In holography, such partition function would correspond to the on-shell value of the gravity action on regular solutions in the Euclidean signature with arbitrary weakly curved boundary metrics. 
To be able to Wick-rotate to Euclidean signature we need time translation invariance and that is also the reason why the gravitational
computation should be restricted to systems in thermal equilibrium as well.

Another example is the effective action approach initiated in \cite{Carter} and revisited more recently in \cite{Dubovsky:2005xd,Dubovsky:2011sj}, which we review in Section~\ref{sec2}. The effective action is given in terms of a set of scalar fields $\phi^I(t,\vec{x})$, which  are usually referred to as Goldstone bosons and can be naturally interpreted as the comoving coordinates of the fluid elements. A key observation is that for featureless incompressible fluids, the only relevant 
property of fluid elements are their volumes. Reparametrizations of the scalars $\phi^I$ which preserve the volume of the
fluid element (i.e. the corresponding Jacobian has unit determinant) are therefore expected to be symmetries of the theory. The requirement
of volume preserving diffeomorphism invariance severely restricts the form of the action. 
It has been shown in \cite{Bhattacharya:2012zx} (see also \cite{Ballesteros:2014sxa}) that the most
general action exhibiting this symmetry is actually unable to capture the most general
dissipationless transport, as obtained from the conventional description. Such discrepancy has triggered generalizations of the above mentioned setup. It has been in fact recently shown in \cite{Haehl:2014zda,Haehl:2015pja} that only after including, among others, a second set of hydrodynamical degrees of freedom analogous to the additional set of fields that one has in the Schwinger-Keldysh formalism \cite{Schwinger:1960qe,Keldysh:1964ud}, the disagreement is no longer there.

In order to understand the relation between the various approaches we decided to set up a precise gravitational dual of fluid effective actions. Our work is very much inspired by \cite{Nickel:2010pr}, as well as \cite{Faulkner:2010jy,Heemskerk:2010hk}. With such a description at hand, one might be able to understand the initial failure of the effective action approach to capture the most general dissipationless fluid and it might allow us to analyze in more detail the separation
between dissipationless and dissipative transport in fluids. In particular, from a gravitational standpoint, it would also allow us to understand to what extent the
dissipation in  fluid/gravity duality is exclusively due to the horizon, and whether the system can be decomposed in a simple 
dissipative near-horizon contribution plus a dissipationless piece, which transmits the near-horizon dynamics to the
ultraviolet (UV) boundary. And, perhaps most importantly, if we
were to understand the relation between gravity and the fluid effective action, we might have a better understanding of why
there exists a fluid/gravity duality in the first place. 

The key step in establishing a relation between fluid effective actions and gravity is to consider
what we call a double Dirichlet problem. This entails fixing a metric on two radial slices in the bulk and performing a computation
of the (partially) on-shell effective action as a function of these two metrics. By construction, the effective action
must be invariant under diffeomorphisms of the two boundary metrics. It is also a consequence of diffeomorphism invariance that
it cannot depend explicitly on the choice of radial position of the two slices. In the absence of additional degrees of
freedom, it is clearly not possible to write a local effective action for the two metrics. This is in agreement with
the fact that the double Dirichlet problem is an underconstrained system. In flat space, for example, it is easy
to construct a linearized gravitational wave which propagates in a spatial direction perpendicular to the ``radial'' direction\footnote{i.e. the direction in which the branes are separated.}
and which does not affect the metrics on two ``radial'' slices. The extra light degrees of freedom that one needs in order to build
a local effective action are a set of Goldstone bosons $\phi^M$, where $M$ runs over both time and space. The reason behind  
the name Goldstone boson is that if we start with two boundary Minkowski metrics,
the effective action should have a Poincar\'e $\times$ Poincar\'e symmetry. But any solution of the gravitational
field equations will connect the two slices and break the symmetry spontaneously to a diagonal Poincar\'e subgroup and thereby
give rise to a set of Goldstone bosons\footnote{Completely analogous reasoning, albeit applied to gauge fields, explains the emergence of pions in the Sakai-Sugimoto model of holographic QCD \cite{Sakai:2004cn}.}. Though this argument clearly does not apply if we start with two arbitrary metrics on the branes, we will nonetheless stick to the term Goldstone bosons.

With these additional light degrees of freedom included, one can write down an effective action $S[g,G,\phi^M]$ where
$g$ is the metric on one of the radial slices which we will refer to as the UV slice, and $G$ the metric on the other
slice which we will refer to as the infrared (IR) slice. This terminology has an embedding in AdS in mind, but part of our discussion
will be general and in particular the effective action should be symmetric under $g \leftrightarrow G$, up to possible
field redefinitions. 

The Goldstone bosons $\phi^M$ can be interpreted as maps from one boundary to the other. In some sense they are bifundamental
fields for the diffeomorphism $\times$ diffeomorphism invariance which the effective action should have, and they
will allow us to pull back the metric on one slice to the other, so that the effective action can be reinterpreted as 
a bigravity theory as in \cite{ArkaniHamed:2002sp}. There has been a lot of interesting work on bigravity theories recently, see, e.g., \cite{Hassan:2011zd}, but that is not a connection which we will be pursuing in this paper.

A geometric construction of the Goldstone bosons proceeds as follows. Given a particular solution of the Einstein's equations
with prescribed metrics on two radial slices, we bring the metric to the radial ADM form $ds^2 = du^2 + g_{\mu\nu}(u,y^{\mu})dy^{\mu}
dy^{\nu}$ with identity lapse and zero shift. In these new coordinates the original radial slices will no longer be
at $u={\rm constant}$ but at $u=u_{1,2}(y^{\mu})$ for some functions $u_{1,2}(y^{\mu})$. Nevertheless, spatial geodesics in the metric
are of the form $y^{\mu}={\rm constant}$ and the Goldstone fields will simply be the map from $(u_2(y^{\mu}),y^{\mu})$ to 
$(u_1(y^{\mu}),y^{\mu})$. By undoing the change of coordinates that put the metric in radial ADM form we obtain the Goldstone
modes in the original variables. By construction, these Goldstone modes are covariant and transform in the right way 
under diffeomorphisms of the metrics on the two slices. We could also imagine alternative 
definitions based on spacelike or null geodesics which make a prescribed angle with one of the two boundaries, 
but expect these to be related through a field redefinition to the previous construction.

In this paper we construct the effective action for conformal fluids from a gravitational embedding in AdS with a horizon in two different ways.
In Section~\ref{sec3} we write down the most general functional to lowest order in a derivative expansion and fix all the freedom by exploiting a suitable subset of solutions to Einstein's equations with double Dirichlet boundary conditions. The resulting leading order covariant effective action is fully nonlinear.

In the second approach developed in Section~\ref{sec4} we construct the linearized effective action order by order in a derivative expansion. This is achieved by explicitly evaluating the (partially) on-shell bulk action on the solutions to Einstein's equations for linearized gravitational perturbations with double Dirichlet boundary conditions on top of an AdS black brane background. Our approach in Section~\ref{sec4} is particularly explicit when it comes to demonstrating the emergence of effective degrees of freedom and, as one might have expected, the result agrees with the construction of Goldstone modes described above. 
In this formalism, it is important to separate
the Einstein's equations into two sets: the Einstein's equations with non-radial indices which need to be solved for and the remaining
equations, which will correspond to the field equations for the Goldstone bosons and should therefore not be solved for. Hence, our effective
action still contains some off-shell information\footnote{Putting the Goldstones on-shell would result in a nonlocal effective action.}.

The effective action for the double Dirichlet problem contains information on how to transmit data from one radial position to another. In fact, computing the on-shell bulk action amounts to ``integrate out'' geometry between the two boundaries and replace it with a simple \emph{local} functional 
$S[g,G,\phi^M]$ living on one of the two slices. From the dual field theory point of view this operation can be interpreted as integrating out high energy degrees of freedom  \'a la Wilson, as in \cite{Faulkner:2010jy,Heemskerk:2010hk}. The resulting boundary effective action acts then as a boundary condition and as a link to the UV for a suitable dynamical (strongly coupled) IR sector extending between the IR brane and the interior of the spacetime in the  
spirit of semi-holography \cite{Faulkner:2010tq}. 
 
In this paper we will consider three different types of IR dynamics. The first naive IR sector that we will consider is
to take the IR Dirichlet boundary condition all the way to the horizon of a black hole. This amounts to imposing Dirichlet
boundary conditions on the horizon itself, which in our effective action corresponds to a degenerate limit of the IR
metric. One may wonder whether this is a physically reasonable thing to do, and as we will see in general it is not.
However, if we work to lowest order in the gradient expansion, this limit makes perfect sense. 
When we take the IR boundary to coincide with the horizon and send the
other boundary to the boundary of AdS, our effective action (after the subtraction of a suitable counterterm) exactly agrees
with the action of a perfect conformal fluid at leading order considered, e.g., in \cite{Dubovsky:2005xd,Dubovsky:2011sj,Bhattacharya:2012zx}. 
Our findings, described in Section~\ref{sec3}, are compatible with volume-preserving diffeomorphism invariance and can be viewed as an alternative ``derivation'' of fluid/gravity duality at least to the order we worked with. We find it however rather intriguing why this should be the case. Why is the low energy dynamics of black branes in AdS compatible with fluid dynamics after all? Why is it not resembling, for example, jelly dynamics, which responds nontrivially to shear stresses too and does not have the above mentioned internal symmetry? Unfortunately we have not been able to answer those questions from  first principles within our gravitational embedding. 

In Section~\ref{sec4} we also consider what happens when we impose a Dirichlet boundary condition on the horizon and expand the theory to higher
order in frequencies and momenta. Although the effective action is still compatible with the volume-preserving diffeomorphism symmetry, we find, 
perhaps not surprisingly, that the answer is in general  divergent, but remains finite when we restrict to stationary configurations. 

To cure these divergences, in Section~\ref{sec5.mp} we consider a second setup, where we couple the effective theory to a dissipative IR system, which
is supposed to describe the near-horizon physics. In principle, one can write down effective actions for the IR as well, but since
the IR describes a finite temperature system this would involve using the Schwinger-Keldysh formalism \cite{Schwinger:1960qe,Keldysh:1964ud} and a doubling of the degrees of freedom. In a gravitational setup this should be realized by a two-sided AdS black hole  \cite{Herzog:2002pc}. Instead, we will couple the effective theory to a simple membrane paradigm  boundary condition \cite{Damour:1978cg,Thorne:1986iy,Iqbal:2008by,Parikh:1997ma} a small distance away from the horizon and
then take the membrane to the horizon. Technically the coupling to the membrane simply modifies the IR boundary condition in
such a way that it imposes infalling, as opposed to Dirichlet, boundary conditions in the IR. There are a few subtleties with
this version of the membrane paradigm that we discussed in our previous paper \cite{deBoer:2014xja}, but here we will see that it correctly
reproduces the dispersion relations of the conformal holographic fluid. 

As the above shows, finding the effective action for the double Dirichlet problem is a useful intermediate step and
clarifies many aspects of the fluid/gravity duality, the interpretation of the Goldstone bosons, and the emergence of volume-preserving
diffeomorphism invariance. It does however not yet provide a clear separation between the dissipative and dissipationless part of
hydrodynamics. Therefore, we also consider a third IR boundary condition in Section~\ref{sec5.epf} by switching to Euclidean signature and imposing 
regularity/smoothness in the IR. This can only be done for stationary configurations, and the IR boundary condition can
be imposed by adding a simple functional of the IR metric to the effective action we obtained before. This functional
captures the contribution of the tip of the Euclidean cigar which describes the Euclidean black hole. If we then extremize
the sum of this IR functional and our double Dirichlet effective action we automatically obtain the lowest order contribution
to the equilibrium partition function considered in \cite{Banerjee:2012iz,Jensen:2012jh}. The extremalization procedure turns out to be equivalent to a Legendre transform which
transforms the energy density into the pressure. This is in perfect agreement with the fact that the action for a fluid in terms
of Goldstone bosons is given by the energy (see Section~\ref{leadingeff}), while the equilibrium partition function is given by the pressure. 

Finally, let us emphasize that the effective action formalism of Ref.~\cite{Dubovsky:2005xd,Dubovsky:2011sj,Bhattacharya:2012zx} is dissipationless by construction, since the entropy current is identically (off-shell) conserved (see Eq.~\eqref{entropyexact}). It is natural to think that this feature should be related to some symmetry of the effective action, as recently discussed in \cite{Haehl:2014zda,Haehl:2015pja}. 
In Section \ref{sec6} we show that there is indeed  a nontrivial transformation of the Goldstones, which, if assumed to be a symmetry of the effective action, correctly reproduces the entropy current as the Noether current of such would-be symmetry. 
 It would be interesting to explore further the connection between these findings and the statements in \cite{Haehl:2014zda,Haehl:2015pja} where the  current related to the adiabaticity equation (an off-shell generalization of the on-shell entropy current conservation) is related to the Noether current of a certain $U(1)$ symmetry appearing in their master Lagrangian. 

Let us briefly summarize our findings. We have constructed, in two different ways, an effective action which captures the low-energy
physics of the double Dirichlet problem in gravity. The relevant degrees of freedom are a set of scalar fields which one can interpret as
Goldstone bosons. In the limit where one of the two boundaries approaches the horizon of a black brane, the
effective action to the lowest order in the derivative expansion becomes that
of a perfect fluid, in agreement with \cite{Dubovsky:2005xd,Dubovsky:2011sj} and the generalization of \cite{Nickel:2010pr}, but at higher orders this near-horizon limit is singular.
By coupling the theory to a Euclidean IR sector we recover the equilibrium partition function of a fluid \cite{Banerjee:2012iz,Jensen:2012jh}, and by coupling to a suitable membrane paradigm boundary condition we obtain dissipative hydrodynamics.

We conclude this Section with the discussion of open questions and new research directions. The double Dirichlet problem can alternatively be interpreted as the transition amplitude of gravity in radial quantization from a
Hartle-Hawking \cite{Hartle:1983ai} point of view. It would be interesting to develop this picture in more detail, and also consider the analogue
problem in de Sitter space, where it could shed further light on the relation between de Sitter correlation functions and
Euclidean partition functions. It is also tempting to use the effective action to find a description of spacetimes with a hole, making contact with the ideas developed in \cite{Balasubramanian:2013rqa,Balasubramanian:2013lsa}.

Besides the IR boundary conditions described in Section~\ref{sec5}, there are other boundary conditions one often encounters, such as
the near-horizon AdS${}_2$ boundary conditions near extremal black branes which feature prominently in various AdS/CMT
applications (see, e.g., \cite{Faulkner:2009wj}). Such strongly coupled IR boundary conditions would, in combination with our effective action, lead to a gravitational version of semi-holography \cite{Faulkner:2010tq} which would be clearly interesting to explore further.

The lowest order effective action in Section~\ref{sec3} was made out of two ``metrics'' and some of our computations are reminiscent
of work done on bi-gravity theories, see, e.g., \cite{Hassan:2011zd}. It is not clear to us whether our work can be used to
come up with holographic duals of certain bi-gravity theories but if it does it might help in understanding their physics.

There are many other directions to explore and we hope to address some of these in the future. This includes, in particular, double Dirichlet problem and corresponding effective actions in the large-D limit \cite{Emparan:2013moa}, an extension of our work to two-sided AdS which naturally gives to the doubled set of degrees of freedom
one needs in the Schwinger-Keldysh formalism to describe dissipation \cite{Herzog:2002pc}; the interpretation of the IR flow in Section~\ref{sec6} which,
in the near-horizon limit, becomes a symmetry whose conserved charge is the entropy; the relation of this analysis to Wald
entropy \cite{Wald:1993nt}; the connection of our work to the
various descriptions of dissipationless fluids which appear in \cite{Haehl:2014zda,Haehl:2015pja};  the study of terms higher
order in the fields and/or derivatives in both gravity and in the effective actions; and possible generalizations to other systems such as solids, superfluids, etc, see, e.g., \cite{Nicolis:2013lma}. Finally, hydrodynamic effective actions appeared recently in a model of dense nuclear matter \cite{Adam:2015rna} and it would be very interesting to pursue this connection further.

{\bf Note added:} While this work was being finalized, we learned that \cite{Crossley:2015tka} will also contain a derivation of the fluid effective action from holography.

{\bf Note added (v2):} The results of \cite{Crossley:2015tka} are in perfect agreement with our approach.

\section{Fluid effective actions: general discussion\label{sec2}}

\subsection{The leading order effective action}\label{leadingeff}

Consider an uncharged relativistic perfect fluid in $d-1$ spatial dimensions. The relativistic version of the Navier-Stokes equations are the conservation equations 
\be
\label{consTmunu}
\nabla_{\mu} T^{\mu \nu} = 0
\ee
of the energy-momentum tensor obeying the following constitutive relation
\be
\label{Tideal}
T^{\mu \nu} = \epsilon \, u^{\mu} u^{\nu} + P \left( g^{\mu \nu} + u^{\mu} u^{\nu} \right).
\ee
This description of an uncharged relativistic fluid utilizes $d$ degrees of freedom: the local energy density $\epsilon$ and the pressure $P$ are related by the equation of state and the fluid velocity $u^{\mu}$ is normalized $g_{\mu \nu} u^{\mu} u^{\nu} = -1$. Fluids described by Eq.~\eqref{Tideal} are perfect, because their evolution does not convert the kinetic and potential energy to heat, i.e. there is no entropy production. This lack of dissipation is captured by the \emph{on-shell} conservation of the entropy current
\be
\label{eqJ}
\nabla_{\mu} \left( s \, u^{\mu} \right) = 0,
\ee
where $s$ is the thermodynamic entropy density expressed in terms of the local energy density or the pressure.

As we have anticipated in the introduction, relativistic perfect fluids admit an alternative description
in terms of the least action principle, see, e.g., \cite{Dubovsky:2005xd,Dubovsky:2011sj,Bhattacharya:2012zx}.
 The structure of this action is such, that the relevant Euler-Lagrange equations carry the same information as the conservation of the corresponding energy-momentum tensor. The relevant Lagrangian turns out to depend on $d-1$ scalar fields
\be\label{dof}
\phi^I=\phi^I(t,\vec{x})\quad\text{where}\quad I=1,\dots,d-1,
\ee
which is less than the number of degrees of freedom in Eq.~\eqref{Tideal}. Perhaps the most natural interpretation of the scalars \eqref{dof} is that of a map at fixed lab-frame time $t$ between space coordinates $\vec{x}$ labelling the Eulerian frame and the internal coordinates $\phi^I$. The latter label the Lagrangian (comoving) frame, i.e.,
$\phi^I(t,\vec{x})$ describes which volume element $\phi^I$ is seen by a fixed Eulerian observer at position $\vec{x}$ when the lab-frame time $t$ is varied. The internal parametrization of the fluid elements is not unique, there is always an obvious freedom of shifting or rotating the fluid elements
\begin{equation}
\label{int1}
\phi^I\rightarrow \phi^I+c^I\quad \mathrm{and} \quad \phi^I\rightarrow R^I_J \,\phi^J. 
\end{equation}
It turns out, however, that the description of perfect fluids requires a much larger symmetry group: invariance under all reparametrizations that do not compress or dilute fluid cells. This is expressed by demanding invariance of the action under the volume-preserving diffeomorphisms in the space of $\phi^{I}$ fields:
\be\label{int2}
\phi^I\rightarrow \xi^I(\phi^{J})\quad\text{with}\quad \det \left(\frac{\partial \xi^I}{\partial\phi^J}\right)=1.
\ee
In particular, this invariance is what mathematically distinguishes between a fluid and a jelly, as in the latter case the internal symmetry reduces to rotational and translational invariance only (\ref{int1}), so that a jelly could respond to shear stresses, see, e.g., \cite{Endlich:2012pz}. In the case of solids, one instead imposes relevant discrete rotational and translational invariance.

The effective action for a relativistic perfect fluid is  then given by
\be\label{eff0}
S=\int d^dx\,\sqrt{-g}\,F(s),
\ee
where
\be
\label{eq.entropystd}
s=s_0\sqrt{\det\partial_{\mu}\phi^I\,\partial^{\mu}\phi^J},
\ee
$s_0$ is a suitable normalization constant and $g$ is the deteminant of the background metric $g_{\mu\nu}$.
The argument $s$ of the yet unspecified scalar function $F$ is proportional to the unique invariant of spacetime and internal symmetries that can be constructed out of the fields $\phi^{I}$ and the background metric $g_{\mu \nu}$ restricting to the lowest possible number of derivatives. 
The combination in the square root of (\ref{eq.entropystd}) is dimensionless given that the fields $\phi^I$ are the comoving coordinates and carry the length dimension. 

The conserved energy-momentum tensor for the action (\ref{eff0}) takes the form
\be\label{Bij}
T_{\mu\nu}=-\frac{2}{\sqrt{-g}}\frac{\delta S}{\delta g^{\mu\nu}}=-sF'(s)B_{IJ}^{-1}\partial_{\mu}\phi^I\partial_{\nu}\phi^J+F(s)g_{\mu\nu} \quad\text{with}\quad B^{IJ}=\partial_{\mu}\phi^I\partial^{\mu}\phi^J
\ee
and becomes the energy-momentum tensor of a perfect fluid \eqref{Tideal} upon identifying $F(s)$ with (minus) the energy density
\be
\epsilon(s)=-F(s)
\ee
and $s$ with the thermodynamic entropy. The entropy current is defined as the spacetime Hodge dual of the volume form in the internal space of the comoving coordinates
\be
\label{entropyexact}
J^{\mu}=s_0\,{}^{*}(d\phi^1\wedge \dots\wedge d\phi^{d-1})=\frac{s_0}{(d-1)!}\epsilon^{\mu\nu_1\dots\nu_{d-1}}\epsilon _{I_1\dots I_{d-1}}\partial_{\nu_1}\phi^{I_1}\dots \partial_{\nu_{d-1}}\phi^{I_{d-1}}
\ee
and is \emph{identically} conserved. The velocity field obtained using the entropy current definition in Eq. \eqref{eqJ} is the same as the one required for the successful identification of the energy-momentum tensors \eqref{Tideal} and \eqref{Bij}.

The equations of motion for $\phi^{I}$ derived from the effective action \eqref{eff0} turn out to be the conservation of the energy-momentum tensor \eqref{Tideal} projected transversally to the flow
\be
\label{consT}
\left( g^{\mu\nu} + u^{\mu} u^{\nu}\right) \nabla^{\rho} T_{\rho \nu}=0.
\ee 
The remaining component of the conservation equation  $u^{\nu}\nabla^{\mu}T_{\mu\nu}=0$, which incorporates the conservation of energy, is implied by the conservation of the entropy current (\ref{eqJ}).

The action \eqref{eff0} receives corrections carrying higher number of derivatives of $\phi^{I}$ fields and such corrections, assuming invariance under the volume-preserving diffeomorphisms \eqref{int2} as the exact symmetry, were obtained up to the second order in the gradient expansion in Ref.~\cite{Bhattacharya:2012zx}. The physics of such fluids is intrinsically non-dissipative as the entropy current is by construction identically (and off-shell) conserved. Let us also mention here that the gravitational calculation in Ref.~\cite{Nickel:2010pr} indicates that including the dissipation requires relaxing the volume-preserving diffeomorphism invariance as the exact symmetry at subleading orders in the gradient expansion. On physical ground, the presence of shear viscosity implies that the fluid would respond nontrivially to shear stresses. This provides an excellent motivation for exploring possible generalizations of the action \eqref{eff0} and its embedding in holography, which are precisely the issues we address in the current work.

\subsection{The linearized expansion}

In the rest of the current Section we assume that the background metric is flat: $g_{\mu \nu} = \eta_{\mu \nu}$. A natural way to fix the fluid's parametrization is requiring that in equilibrium and on a given time slice the fluid elements are aligned with the spatial coordinates  
\be\label{groundstate}
\phi^I(\vec{x},t)=\delta^I_{\mu}x^{\mu}.
\ee
This configuration spontaneously breaks the spacetime Poincar\'e symmetry and the global subgroup of the internal symmetry (\ref{int1}-\ref{int2}) down to diagonal rotations and spatial translations. Although there should be one Goldstone boson per broken generator, it turns out that in the presence of spacetime symmetries not all the Goldstones are independent, see, e.g., \cite{Low:2001bw}. It has been shown in \cite{Nicolis:2013lma}, by means of the coset construction, that the only independent Goldstone bosons correspond to the breaking of the space and internal translations down to the diagonal combination of thereof. At the linearized level, such Goldstones are realized as perturbations on top of the equilibrium configuration
 (\ref{groundstate})
\be
\label{eq.goldstones}
\phi^i(\vec{x},t)=x^{i}+\pi^i(t,\vec{x}),\quad i=1,\dots,d-1.
\ee
In the formula above, we do not distinguish the internal and coordinate indices since the Goldstone bosons transform under the diagonal combination thereof.

Let us consider now linearizing the fluid's action in the Goldstone fields \eqref{eq.goldstones}. The Goldstones can be classified according to their orientation with respect to the propagation direction being longitudinal or transverse
\be
 \vec{\pi}=\vec{\pi}^L+\vec{\pi}^T\qquad \text{with}\quad \vec{\nabla}\times \vec{\pi}^L=0\quad \text{and}\quad \vec{\nabla}\cdot \vec{\pi}^T=0.
\ee
Linearization of the velocity field and the entropy density give
\ba
u^t=-1-\frac{1}{2}(\partial_t\vec{\pi})^2+\dots,\quad \vec{u}=\partial_t \vec{\pi}+\dots\\
s=s_0+s_0\nabla\cdot\vec{\pi}-\frac{1}{2}s_0(\partial_t\vec{\pi})^2+\dots
\ea
and the effective action (\ref{eff0}) becomes 
\be
S^{(0)}=\int d^d x\left\{F(s_0)-\frac{1}{2}F'(s_0)s_0\Big((\partial_t\vec{\pi}^T)^2+(\partial_t\vec{\pi}^L)^2-c_s^2(\nabla\cdot \vec{\pi}^L)^2\Big)+\dots\right\}. 
\ee
The equations of motion, equivalent to the conservation \eqref{consT} of the energy-momentum tensor \eqref{Tideal}, give the following leading order dispersion relations
\ba
\pi^L: \quad \omega_L&=&\pm c_s k,\\
\pi^T: \quad \omega_T&=&0\label{sheardisp}.
\ea
The longitudinal Goldstone describes a sound wave. Its group velocity $c_s$ is given by
\be
c_{s}^2=\frac{F''(s_0)}{F'(s_0)}s_0=\frac{P'(s_0)}{\epsilon'(s_0)}=\frac{dP}{d\epsilon}.
\ee
For a conformal fluid
\be
F(s) \sim s^{d/d-1}
\ee 
and the speed of sound takes the familiar form
\be
\label{eq.speedofsound}
c_s=\frac{1}{\sqrt{d\!-\!1}}.
\ee

The transverse Goldstones do not propagate, because their gradients do not contribute to the action (energy). This is a direct consequence of the volume preserving diffeomorphism invariance of the action \eqref{eff0}. At a linearized level, the volume preserving diffeomorphisms act as follows
\be
\vec{\pi}(t,\vec{x})\rightarrow \vec{\pi}(t,\vec{x})+\vec{\xi}(\vec{x})\quad \text{with }\quad \nabla \cdot \vec{\xi}=0
\ee
and do not allow the gradient terms of the form $\nabla \times \vec{\pi}$ to appear in the action.

\subsection{General internal space metric and timelike Goldstone}\label{TimeGold}

Let us now consider two generalizations of the perfect fluid action \eqref{eff0} which are motivated by the holographic correspondence and the results of Ref. \cite{Nickel:2010pr}. Perhaps the most natural generalization of the previous construction is a theory with arbitrary, albeit nondynamical, metric in the configuration space $G_{ij}(\vec{\phi})$
\be
s=s_0\sqrt{\det \left( \partial_{\mu}\phi^i\,\partial^{\mu}\phi^j\,G_{jk} \right)}.
\ee
The entropy current \eqref{entropyexact} depends now on the configuration space metric via the Levi-Civita tensor
\be
\epsilon_{i_1\dots i_{d-1}}\rightarrow \sqrt{\det G_{ij}}\,\epsilon_{i_1\dots i_{d-1}}^{(0)},
\ee
where $\epsilon^{(0)}$ is the (flat space) Levi-Civita symbol. Nevertheless, it is still identically conserved
\be
\nabla _{\mu}J^{\mu}=-\frac{1}{2}J^{\mu}(\partial_{\mu}G_{ij})G^{ij}=-\frac{1}{2}J^{\mu}(\partial_{\mu} \phi^k)(\partial_k G_{ij})G^{ij}=0,
\ee
as $\phi^{i}$ fields are the comoving coordinates and hence $J^{\mu}\partial_{\mu}\phi^i = s \cdot (u^{\mu} \partial_{\mu} \phi^{i})=0$.

A more complicated generalization follows from considering systems parametrized by $d$, instead of $d-1$, scalar fields. At least superficially, one might think of such systems as containing also a timelike Goldstone boson as a low energy excitation of the ground state:
\be\label{Goldlin}
\phi^{M}(x^{\mu})=\delta^{M}_{\mu}x^{\mu}+\pi^{\mu}(x^{\mu}),
\ee
where now $\phi^{M}=(\phi^t,\vec{\phi})$. Having in mind a gravitational embedding we restrict to the case in which the metric in the internal space is degenerate. A convenient parametrization of such metric is given in terms of the Galilean metric \cite{Nickel:2010pr}, which is defined as
\be\label{IRmetric}
ds^2=G_{MN}d\phi^Nd\phi^M=G_{ij}(d\phi^i-v^id \phi^t)(d\phi^j-v^jd \phi^t),
\ee
and the following null vector 
\be
 n^M=\frac{1}{\gamma}(1,v^i) \quad \text{with}\quad G_{MN}n^M=0\quad \text{and}\quad\gamma=(1-G_{ij}v^iv^j)^{-1/2}.
\ee
The fields $G_{ij}$ and $\vec{v}$ depend on the internal space coordinates $\phi^M$. 

The entropy current \eqref{entropyexact} can be generalized to be the spacetime Hodge dual of the volume form of the boosted coordinates $ e^i$
\be
\label{eq.genentropycurrent}
J^{\mu}=\frac{s_0}{(d-1)!}\,\epsilon^{\mu\nu_1\dots\nu_{d-1}}\epsilon _{i_1\dots i_{d-1}}e_{\nu_1}^{i_1}\dots e_{\nu_{d-1}}^{i_{d-1}},
\ee
where the combinations $e_{\mu}^i$ are 
\be
e_{\mu}^i=\partial_{\mu}\phi^i-v^i \partial_{\mu}\phi^t.
\ee
The Greek indices $\mu$, $\nu$, etc. are raised with the spacetime inverse metric $g^{\mu\nu}$ and the Latin indices $i$, $j$, etc. are lowered and raised with the configuration space metric $G_{ij}$ and its inverse. Writing explicitly the dependence of the Levi Civita tensors on the relevant metrics, the generalized entropy current takes the form
\be\label{jnew}
J_{\mu}=\frac{s_0}{(d-1)!}\sqrt{\det G_{ij}}\sqrt{-g}\,g^{\nu_1\lambda_1}\dots g^{\nu_{d-1}\lambda_{d-1}}\,\epsilon^{(0)}_{\mu\lambda_1\dots\lambda_{d-1}}\epsilon^{(0)}_{i_1\dots i_{d-1}}e_{\nu_1}^{i_1}\dots e_{\nu_{d-1}}^{i_{d-1}},
\ee
and the corresponding entropy density is
\be\label{entropynew}
s=s_0\sqrt{\det (e_{\mu}^ie^{\mu\,j}\,G_{jk})}.
\ee
Note that a priori the current \eqref{eq.genentropycurrent} is \emph{not} conserved off-shell for generic $G_{ij}$ and $\vec{v}$.

Let us now demonstrate that both generalizations of the conventional effective field theory of fluids exposed in Section \ref{leadingeff} lead to the correct equations of relativistic fluid mechanics. Consider for concreteness the following configuration  
\be\label{metricIR}
G_{ij}=\delta_{ij}+H_{ij}(\phi^M);\qquad v_i=-H_{ti}(\phi^M),
\ee
where $H_{i j}$ and $H_{t i}$ are small. Without any loss of generality we can restrict the dependence of the perturbations $H$ to $(\phi^t,\phi^x)$ and divide the perturbations according to their transformation properties in the remaining transverse $O(d-2)$ plane
\ba
\text{scalar:}&\quad& H_{xx},\quad H,\quad H_{tx},\quad  H_{tt}\nn \\
\text{vector:}&\quad& H_{x\alpha},\quad H_{t\alpha}\nn\\
\text{tensor:}&\quad& H_{\alpha \beta}-\frac{1}{(d-2)}\delta_{\alpha \beta} H,
\ea
where $H=\sum_{\alpha} H_{\alpha\alpha}$ is the trace in the transverse direction $\alpha = 1,\dots d-2$. 

Inserting the scalar and vector sectors expansions into (\ref{entropynew}) and using the linearized Goldstone expansions (\ref{Goldlin}), the leading order effective action (\ref{eff0}) takes the form
\ba\label{effactionfull}
S^{(0)}&=&\int d^dx\Bigg(F(s_0)+\frac{1}{2}s_0F'(s_0)H_{ii}+s_0F'(s_0)\d_x\pi^x+\nn\\
&&\qquad -\frac{1}{2}s_0F'(s_0)\left(H_{tx}^2+\frac{1}{4}H_{xx}^2-\frac{1}{2}H_{xx}H-\frac{1}{4}c_s^2H_{ii}^2\right)+\nonumber\\
&&\qquad +s_0F'(s_0) \Big(\frac{1}{2}\dot{H}_{ii}-\partial_x H_{tx}\Big)\pi^t+\nonumber\\
&&\qquad-\frac{1}{2}s_0F'(s_0)\Big((\dot{\pi}^x)^2-c_s^2(\d_x \pi^x)^2-2\,\pi^x\dot{ H}_{tx}+c_s^2\,\pi^x\,\partial_x H_{ii}\Big)+\nn\\
&&\qquad -\frac{1}{2}s_0F'(s_0)\sum_{\alpha}\Big((\dot{\pi}^{\alpha})^2+H_{t\alpha}^2+ H_{x\alpha}^2 -2\,\pi^{\alpha}\dot{H}_{t\alpha}\Big) \Bigg),
\ea
where $H_{ii}=H_{xx}+H$.
 Notice that the timelike Goldstone $\pi^t$ appears  as a Lagrange multiplier and the corresponding equation of motion ensures now the on-shell conservation of the entropy current
\be\label{Jconsnew}
\nabla _{\mu}J^{\mu}= -\frac{1}{2}\dot{H}_{ii}+\partial_x H_{tx}+\dots \big|_{\text{on-shell}}=0.
\ee
  The equations of motion for the transverse and longitudinal Goldstones are now respectively
\be
\partial_t^2\pi^{\alpha}+\partial_t H_{t\alpha}=0,\quad \quad\partial_t^2\pi^x-c_s^2\,\partial_x^2 \pi^x+\, \partial_t H_{tx}-\frac{1}{2}c_s^2\,\partial_xH_{ii}=0
\ee
and correspond to the conservation equations (\ref{consT}) of the perfect fluid stress energy tensor 
 (\ref{Tideal}) with the velocity and the entropy density \emph{redefined} in the following way
 \ba
u^t=-1-\frac{1}{2}(H_{t\vec{x}}+\partial_t\vec{\pi})^2+\dots,\quad \vec{u}=H_{t\vec{x}}+\partial_t \vec{\pi}+\dots\\
s=s_0+\frac{1}{2}s_0\, H_{ii}+s_0\nabla\cdot\vec{\pi}+\dots. \qquad \qquad
\ea
These expressions can be obtained from the linearization of (\ref{entropynew}) and using $u^{\mu}=J^{\mu}/s$, where $J^{\mu}$ given in (\ref{jnew}). Hence we showed explicitely that the above generalizations reproduce the correct hydrodynamic equations. It can be  shown that the same is true at nonlinear level.
This analysis indicates that the standard action for relativistic perfect fluids is a particular instance of the more general action obtained by making the Lagrangian \eqref{eff0} depend on $s$ defined by Eq.~\eqref{entropynew}, rather than by Eq.~\eqref{eq.entropystd}. Actions of these types were encountered previously in Ref.~\cite{Nickel:2010pr} and in the following sections we will derive such an action using holography. Notice that a key ingredient in this derivation is the degenerate nature of the configuration space metric. Had we worked with a general non-degenerate metric instead, we would have had an additional dynamical but nonhydrodynamic degree of freedom: the timelike Goldstone. 

\section{Fluid effective action: holographic derivation\label{sec3}}

Inspired by the deconstruction of holographic fluids proposed in \cite{Nickel:2010pr}, we are interested in obtaining the effective action
\be\label{effaction}
S_{\rm eff}(g_{\mu\nu},G_{MN},\phi^M),
\ee
invariant under diffeomorphisms on two branes endowed with two independent metrics $g_{\mu\nu}$ and $G_{MN}$. The Goldstone modes $\phi^M(x^{\mu})$ are the bifundamental fields between two such theories of gravity and can be used to rewrite the effective action (\ref{effaction}) as a 
\emph{local} theory on one of the branes
\be\label{effaction1}
S_{\rm eff} = \int d^d x \, \sqrt{-g}\, F(g_{\mu\nu},h_{\mu\nu}),
\ee
where $h_{\mu\nu}$ is the pull-back of the metric $G_{MN}$ on one of the branes to the other brane
\be
\label{eq.defh}
h_{\mu\nu} = G_{MN} \frac{\partial \phi^M}{\partial x^{\mu}} 
\frac{\partial \phi^N}{\partial x^{\nu}}
\ee
and $F(g_{\mu\nu},h_{\mu\nu})$ is a scalar quantity built from the two tensors $g$ and $h$.

In principle, the action \eqref{effaction} could contain higher derivative terms, e.g. second derivatives of the Goldstone fields or the curvatures built from $g$ and $h$. Here, we are interested in the lowest order terms only, i.e.
terms with arbitrary numbers $k$ of Goldstone fields and $l\leq k$ derivatives. This is the analogue of the gradient expansion from the previous Section and is also the reason why we assume the effective action depends only on the metrics and the Goldstones.

The most general scalar we can construct from $g_{\mu\nu}$ and $h_{\mu\nu}$, without using derivatives, is a function of the traces $ {\rm Tr}(M^k)$ of the matrix
 $M\equiv h^{\mu \rho} g_{\rho \nu}$, where $h^{\mu\rho}$ is the inverse of $h_{\mu\rho}$. For $d\times d$ dimensional matrices there are only $d$ independent traces, corresponding to the amount of eigenvalues. The effective action (\ref{effaction}) will then, in general, depend on $d$ scalars through 
\be\label{scalaraction}
F(g_{\mu\nu},h_{\mu\nu})=F[M]=F[{\rm Tr}(M),\,\dots \,,{\rm Tr}(M^d)].
\ee
Equivalently, we could have chosen to work with traces of $M^{-1}$, but will find traces of $M$ to be
more convenient.

\subsection{The double Dirichlet problem and derivation}

So far the discussion has been rather general. However we can be more specific and think of the effective action~(\ref{effaction}) as embedded in a higher dimensional spacetime where the additional coordinate plays the role of an energy scale. Going deeper in the interior of spacetime would correspond to approaching the low energy regime of the theory. Such scenario is the one that is precisely realized in holography, where the two branes can be thought of as being located on two finite cutoffs and can therefore be dubbed as  IR and UV brane endowed with, respectively, $G_{MN}$ and $g_{\mu\nu}$ metrics.
The low energy effective action can be derived following the holographic Wilsonian renormalization group flow procedure \cite{Faulkner:2010jy,Heemskerk:2010hk}. Focussing on the specific case of asymptotically AdS black-brane background, we will solve the double Dirichlet problem between the two branes and compute the (partially) on-shell action\footnote{Not all Einstein's equations will be used  to compute such action. Constraint equations will be left unsolved and  will turn out to be equations of motion for the Goldstones.} in that region. 
 In this Section we will consider the full nonlinear setup and derive the leading order effective action by looking at a suitable subset of solutions to Einstein's equations. By sending the  
 the UV brane to conformal infinity and the IR brane to the horizon we will reproduce the effective action for conformal fluids (\ref{eff0}).

Let us then consider the general action for Einstein gravity
\be
\label{eq.gravaction}
S=\int du \,d^dx \, \sqrt{-g} \,(R-2\Lambda),
\ee
and the corresponding field equations
\be\label{Einsteq}
R_{ab}=\frac{2\Lambda}{d-1} g_{ab},
\ee
where $a,b = 1,\dots d+1$ are spacetime indices. In order to shorten our formulas in this Section, starting from Eq.~\eqref{eq.gravaction} above, we dropped their dependence on the Newton's constant.

Let us restrict to the case where the
metrics on the IR and UV brane are constant. After all, we are interested in the effective action which does not
include derivatives on the IR and/or UV metrics. The most general metric in which the Goldstones, obtained from spatial geodesics (see Introduction), are given by $\phi^M = \delta^M_{\mu} x^{\mu}$ is
\be \label{newform3}
ds^2 = dU^2 + 2 \,A_{\mu}(y^{\nu}) dy^{\mu}dU  + g_{\mu\nu}(y^{\mu},u) dy^{\mu} dy^{\nu},
\ee
where $U$ is an arbitrary function of $u,y^{\mu}$. Because we are assuming that the UV and IR metric do not
depend on $y^{\mu}$, we can take the entire metric to be independent of $y^{\mu}$. We might
then as well use $U$ as our radial variable, which (after we relabel $U$ by $u$) results in 
\be \label{newform4}
ds^2 = du^2 + 2 \,A_{\mu} dy^{\mu}du + g_{\mu\nu}(u) dy^{\mu} dy^{\nu}.
\ee
The following shift $y^{\mu} \rightarrow y^{\mu} -u \,A^{\mu}$ gets rid of $A_{\mu}$ from the metric. The Ricci
scalar as a function of the single variable $g$ can then be evaluated to be
\be
R = -{\rm Tr} (g^{-1}\partial_u^2 g) +\frac{3}{4} {\rm Tr}(g^{-1}\partial_u g\, g^{-1}\partial_u g)
-\frac{1}{4} ({\rm Tr}(g^{-1}\partial_u g))^2,
\ee
with
\bea
R_{uu} & = & -\frac{1}{2} {\rm Tr} (g^{-1}\partial_u^2 g) +\frac{1}{4}{\rm Tr}(g^{-1}\partial_u g\, g^{-1} \partial_u g), \\
R_{\sigma\nu} & = & -\frac{1}{2} \partial_u^2 g + \frac{1}{2} \partial_u g \,g^{-1} \partial_u g -\frac{1}{4}
{\rm Tr}(g^{-1}\partial_u g) \partial_u g   \label{eq2}.
\eea
We
anticipate that the effective action only involves the eigenvalues of the matrix~$M$ through (\ref{scalaraction}), and we can probe this already with diagonal
metrics on the two boundaries. Because of this, we will now restrict to diagonal metrics. 

If we multiply (\ref{eq2}) with the inverse of $g$, take the trace and use the Einstein's equations (\ref{Einsteq}) we obtain
\be\label{eqeinst1}
-\frac{1}{2} \partial_u {\rm Tr}(g^{-1}\partial_u g)  -\frac{1}{4}
({\rm Tr}(g^{-1}\partial_u g))^2 = 2\Lambda \frac{d}{d-1},
\ee
which is a first order equation for the combination ${\rm Tr}(g^{-1}\partial_u g)$. If we assume 
negative cosmological constant and define
\be
\ell^2 \equiv - 2\Lambda \frac{d}{d-1},
\ee
the solution to (\ref{eqeinst1})  is
\be
\label{eq.trg1mgu}
{\rm Tr}(g^{-1}\partial_u g) = 2\ell \coth\ell(u-u_0),
\ee
where $u_{0}$ is an integration constant. Since the left hand side of Eq.~\eqref{eq.trg1mgu} is equal to $\partial_u \log \det g$, we can compute the 
determinant of $g$, which equals
\be
\label{eq.detg}
\det g = C \sinh^2 \ell (u-u_0) 
\ee
with some integration constant $C$. 

Let us now turn back to  Einstein's equations involving (\ref{eq2}). Take a diagonal $g$ and consider one
of its diagonal components which can be parametrized with $\exp{\left( \phi_{\mu} \right)}$. For the time direction there should be a minus
sign but we can ignore this, since all metrics we consider can be Wick rotated so we might as well
work in the Euclidean signature.

Suppressing the index on $\phi$, from Einstein's equations (\ref{Einsteq}) and (\ref{eq2}) we obtain the equation
\be
-\frac{1}{2} \partial_u^2 \phi - \frac{\ell}{2} \tanh\ell(u-u_0)\, \partial_u\phi = -\frac{\ell^2}{d}.
\ee
Its solution is
\be\label{solphi}
\phi_{\mu} = \frac{2}{d}\log\sinh \ell (u-u_0) + A_{\mu} + B_{\mu} \log \tanh \frac{\ell}{2}(u-u_0)
\ee
where $A_{\mu}$ and $B_{\mu}$ are some integration constants. Self-consistency with the equation for the determinant \eqref{eq.detg}
imposes no constraints on $A_{\mu}$, but requires 
\be
\sum_{\mu} B_{\mu}=0.
\ee
Thus, the total number of integration constants seems to be equal to $2d$: $d$ from the $A_{\mu}$, $d-1$ from the $B_{\mu}$, and
one from $u_0$. Naively, this is the right number of integration constants to allow for arbitrary diagonal metrics on the
IR and UV brane.

However, we also need to analyze the remaining $uu$ component of the Einstein's equation:
\be
R_{uu} = \sum_{\mu} \left( -\frac{1}{2} \partial_u^2 \phi_{\mu} - \frac{1}{4} \partial_u \phi_{\mu}\,
\partial_u \phi_{\mu} \right) .
\ee
This equation leads to one more constraint
\be
\sum_{\mu} B_{\mu}^2 = 4\frac{d-1}{d}.
\ee
Do we have sufficiently many integration constants to get arbitrary metrics on the IR and UV branes? Yes, because
we have the freedom to choose the values of $u=u_1$ and $u=u_2$ where the IR and UV brane live\footnote{This is reminiscent of the situation encountered when calculating transition amplitudes in quantum gravity \cite{Hartle:1983ai}.}. The total number
of variables is therefore 
$d$
from $A_{\mu}$, 
$(d-2)$ from $B_{\mu}$, plus $u_0, u_1, u_2$. However, shifting $u_0,u_1,u_2$
simultaneously by a constant does not change the solution, so there are only two independent variables among
$u_0,u_1,u_2$. The total number of free variables is therefore $2 d$, which is precisely the right number.

Let us do a quick sanity check and take $d=2$. We then find that $\sum B_{\mu}^2=2$ and thus, e.g., $B_t=+1$ and $B_x=-1$.
Plugging these values in, we get
\be
\phi_t = \log \sinh^2 \frac{\ell}{2}(u-u_0) + {\rm const},\qquad
\phi_x = \log \cosh^2 \frac{\ell}{2}(u-u_0) + {\rm const},
\ee
and this indeed agrees with the AdS${}_3$ metric of the form
\be
ds^2 = du^2 - \sinh^2 \frac{\ell}{2}(u-u_0) dt^2 + \cosh^2 \frac{\ell}{2}(u-u_0) dx^2 .
\ee

Suppose that we have a diagonal metric ${\rm diag}(-e^{\phi^1_t},e^{\phi^1_x},\ldots)$ at the IR brane at $u=u_1$,
and similarly a diagonal metric ${\rm diag}(-e^{\phi^2_t},e^{\phi^2_x},\ldots)$ at the UV brane $u=u_2$.
We expect the effective action to only depend on the ratio of the IR and UV metric, as that is what appears in the 
matrix $M$ we used above. Indeed, the shift variables $A_{\mu}$ do not appear in the solutions in a very profound way
and do not affect the ratio of the IR and UV metric. In other words, they effectively decouple, as expected.
We are left with the following system of equations  (we set $u_0=0$ for simplicity)
\bea \label{finaleqs}
\phi^2_{\mu} - \phi^1_{\mu} & = & \frac{2}{d} \log\left( \frac{\sinh \ell u_2}{\sinh \ell u_1}\right)
+ B_{\mu} \log\left( \frac{\tanh \frac{\ell}{2} u_2}{\tanh \frac{\ell}{2} u_1}\right), \\
\sum_{\mu} B_{\mu} & = & 0, \\
\sum_{\mu} B_{\mu}^2 & = &  4 \frac{d-1}{d},
\eea
which we need to solve. We can easily solve the first equation for $B_{\mu}$ and are then left with two equations for $u_1$ and $u_2$ which are not particularly easy to solve.

We are now ready to evaluate  the on-shell action which contains besides (\ref{eq.gravaction}) two Gibbons-Hawking terms on the two boundaries.
The Hilbert-Einstein contribution  because the Ricci scalar is constant\footnote{This is certainly true when both the constraints and the dynamical components of the Einstein's equations are imposed. Here, we do not want to impose the constraints associated with the choice of the shift vector $A_{\mu}$. However, in the radial gauge, the relevant off-diagonal contributions from the equations of motion to the Ricci scalar vanish, as the inverse metric is diagonal. This is the reason why also in our setup the Ricci scalar is constant.} reads
\be
S= V \,\frac{4\Lambda}{d-1} \int_{u_1}^{u_2} du \,\exp\left(\sum_{\mu} \phi_{\mu}(u)/2\right),
\ee
where $V$ is the volume in the $t,\vec{x}$ directions. Such action can be easily evaluated using our solution (\ref{solphi}) and takes the final form
\be
S = -\frac{2V\ell}{d} \,e^{\sum_{\mu} A_{\mu}/2}\, (\cosh\ell u_2 - \cosh\ell u_1) .
\ee
The Gibbons-Hawking contribution is of the form 
\be
 S_{GH}=- \int d^d x \, \sqrt{-g} \,{\rm Tr}(g^{-1}\partial_u g),
\ee
and on, e.g., $u_2$ it evaluates to
\be
S_{GH} = -2 V \ell \,e^{\sum_{\mu} A_{\mu}/2} \cosh\ell u_2.
\ee
The final result, combining all three contributions, thus reads
\be \label{f1}
S_{\rm total} = -\frac{2V(d+1) \ell}{d} \, e^{\sum_{\mu} A_{\mu}/2} \, (\cosh\ell u_2 - \cosh\ell u_1) .
\ee
From (\ref{finaleqs}) we obtain that
\bea \label{f2}
\log\left( \frac{\sinh \ell u_2}{\sinh \ell u_1}\right) & = & \frac{1}{2} \sum_{\mu} (\phi^2_{\mu}-\phi^1_{\mu}), \\ \label{f3}
\left(\log\left( \frac{\tanh \frac{\ell}{2} u_2}{\tanh \frac{\ell}{2} u_1}\right)\right)^2 & = & 
\frac{d}{4(d-1)}\left( \sum_{\mu} (\phi^2_{\mu}-\phi^1_{\mu})^2 -\frac{1}{d}
\bigg(\sum_{\mu} (\phi^2_{\mu}-\phi^1_{\mu})\bigg)^2 \right),
\eea
and equations (\ref{f1}), (\ref{f2}) and (\ref{f3}) are the final set of equations we would like to solve.

To compare with the effective action  which we introduced above (\ref{effaction1}) with (\ref{scalaraction})  
we need to insert the value of $\sqrt{-g}$ at the UV brane. We read off that
\be \label{f4}
F[M] = -\frac{2(d+1)\ell}{d} \frac{\cosh\ell u_2 - \cosh\ell u_1}{\sinh \ell u_2}.
\ee
The eigenvalues of $M$ are $\exp(\phi_{\mu}^2-\phi_{\mu}^1)$, i.e. the eigenvalues of the UV metric times the ones of the inverse IR metric.
We have therefore succeeded in writing $F[M]$ in terms of the eigenvalues of the matrix $M$: one first needs to solve for 
$u_1$ and $u_2$ in terms of the eigenvalues using equations (\ref{f2}) and (\ref{f3}) and substitute those in (\ref{f4})
to get the expression of $F[M]$ in terms of its eigenvalues.

One conclusion we can already draw is that
\be
F[M]\equiv F[ {\rm Tr}(\log M), {\rm Tr}((\log M)^2)],
\ee
since those are the only combinations of eigenvalues that appear. And, to summarize once more, the function with 
two arguments that appears on the right hand side is given by
\be\label{effactiont}
F[t_1,t_2] = -\frac{2(d+1)\ell}{d} \frac{\cosh\ell u_2 - \cosh\ell u_1}{\sinh \ell u_2},
\ee
where the relation between $t_1,t_2$ and $u_1,u_2$ is given by
\bea \label{f5}
\log\left( \frac{\sinh \ell u_2}{\sinh \ell u_1}\right) & = & \frac{t_1}{2}, \\ \label{f6}
\left(\log\left( \frac{\tanh \frac{\ell}{2} u_2}{\tanh \frac{\ell}{2} u_1}\right)\right)^2 & = & 
\frac{d}{4(d-1)}\left( t_2 -\frac{t_1^2}{d} \right).
\eea
It appears difficult to obtain the solution for $u_1$ and $u_2$ in any compact form, so this is as close as it gets to finding
an explicit expression for the effective action.

\subsection{Taking the IR brane to the horizon}

The near-horizon limit in our setup is $u_1\rightarrow 0$ with $u_2$ kept finite and it
corresponds to sending $G_{tt} \rightarrow 0$. At the same time, $t_1\rightarrow -\infty$ while $t_2\rightarrow +\infty$. As $u_1\rightarrow 0$ we
find that
\be \label{p1}
\frac{t_1}{2} + \left[ \frac{d}{4(d-1)}\left( t_2 -\frac{t_1^2}{d} \right) \right]^{\frac{1}{2}} = 
2\log\cosh \frac{\ell u_2}{2}
\ee
and the effective action (\ref{effactiont}) reduces to 
\be
F[t_1,t_2] = -\frac{2(d+1)\ell}{d} \tanh \frac{\ell u_2}{2}.
\ee
From Eq.~(\ref{p1}) we see that by degenerating one eigenvalue on the left hand side, this equation reduces to 
\be
2\log\cosh \frac{\ell u_2}{2} = \frac{d}{2(d-1)} {\rm Tr'}\log M,
\ee
where ${\rm Tr'}$ is the trace with the degenerate eigenvalue removed.
We can now solve for $u_2$ and plug it to the effective action. The result is
\be \label{aux101}
F[M]= -\frac{2(d+1)\ell}{d} \left[ 1- \exp \left( -\frac{d}{2(d-1)} {\rm Tr'} \log M \right) \right]^{\frac{1}{2}},
\ee
which is a very concrete effective action. Notice that it only depends on ${\rm Tr'}\log M=\log \det' M$ and, therefore, it is invariant under the volume-preserving diffeomorphisms.

To make the connection to the results in Section~2 more explicit, we work out the explicit form of ${\rm Tr'}\log M$,
assuming the IR metric is of the form $G_{MN}d\phi^M d\phi^N = G_{tt}d\phi^t d\phi^t + G_{ij} d\phi^i d\phi^j$.
In the near-horizon limit, with $G_{tt}\rightarrow 0$, one of the eigenvalues of $M$ will blow up, or equivalently,
one of the eigenvalues of $M^{-1}$ will go to zero. It is then very easy to see that
\be \label{aux102}
\det{} ' M^{-1} = \det\left( \frac{\partial \phi^i}{\partial x^{\mu}} \frac{\partial \phi^j}{\partial x^{\nu}} g^{\mu\nu} G_{jk}
\right)=(s/s_0)^2
\ee
and therefore our action (\ref{aux101}) is indeed of the type (\ref{eff0}).

One could also have rewritten $\det' M^{-1}$ somewhat more covariantly as a function of powers of traces of
$M$, as was done in \cite{Nickel:2010pr}. If, for example, $M^{-1}$ is a $3\times 3$ matrix with eigenvalues
$\lambda_1,\lambda_2,\lambda_3$, then $\lambda_1 \lambda_2 +\lambda_2 \lambda_3 + \lambda_3\lambda_1$ reduces
to the product of the non-zero eigenvalues in case one of the eigenvalues is equal to zero. Moreover, this
symmetric polynomial can be written in terms of powers of traces of $M^{-1}$, and in this way one does need
to introduce the prime notation. We have not done this because such a rewriting depends in a complicated
way on the dimension of spacetime, and moreover our construction utilizing the double Dirichlet problem does not give rise to this structure
if the IR boundary does not coincide with the event horizon of a black brane.

\subsection{Relation to action for conformal perfect fluids}

Finally, let us take the limit where the UV metric blows up (going near the boundary of AdS). Then the exponent
in Eq.~\eqref{aux101} becomes very small and we can approximate
\be
F[M]\approx -\frac{2(d+1)\ell}{d} \left[ 1- \frac{1}{2}\exp \left( -\frac{d}{2(d-1)} {\rm Tr'} \log M \right) \right].
\ee
The first term is a constant, so it can be canceled by a local counterterm proportional to $\int d^d x \sqrt{-g}$,
and the only thing that remains is the second term. The effective action therefore becomes
\be
F[M] = \frac{(d+1)\ell}{d} (\det {}' M)^{-\frac{d}{2(d-1)}}
\ee
and in view of (\ref{aux102}) this is exactly the power that we need to describe a conformal fluid in $d$ spacetime dimensions. 
We have therefore found a direct derivation of the effective action for ideal conformal fluids from holography.

\section{Linearized conformal fluid effective action: from gravity\label{sec4}}

We will now move to the explicit construction of the effective action for the (3+1)-dimensional conformal fluid described by ${\cal N} = 4$ super Yang-Mills in the large-$N_{c}$ limit and at strong coupling. We will focus our attention on the regime where deviations from equilibrium are not only long-wavelength, but also small in their amplitude. This will allow us to extend the analysis from the previous Section to higher orders in the low momentum/frequency expansion. The relevant gravity action is
\be\label{HE}
S=\frac{1}{2k_5^2}\int du\, d^4x\sqrt{-g} \,(R-2\Lambda)
\ee
and the corresponding equations of motion are
\be
E_{ab}=R_{ab} - \frac{1}{2} R\, g_{ab} + \Lambda \, g_{ab} = 0.
\ee
The black brane geometry dual to the plasma state of ${\cal N} = 4$ super Yang-Mills takes the following form
\be\label{bb}
ds^2=\frac{(\pi T L)^2}{u}(-f(u)dt^2+dx^2+dy^2+dz^2)+\frac{L^2du^2}{4u^2f(u)},
\ee
where $u$ is the radial coordinate extending from $u = 0$ (UV boundary) to $u = 1$ (the event horizon), the emblackening factor reads $f(u)=1-u^2$, $T$ is the Hawking temperature and $L$ is the curvature radius of the vacuum AdS$_{5}$. In this convention, $\Lambda = - \frac{6}{L^{2}}$.

In studying small perturbations $\delta h_{ab}(t,x,u)$ of the black brane background~(\ref{bb}) it will be convenient to work in the Fourier space
\be
\delta h_{ab}(t,x,u)=\int \frac{d\omega\, dk}{(2\pi)^2}\,\delta h_{ab}(\omega,k,u)\,e^{-i\omega t+ikx}
\ee
and to further define
\be
\label{eq.perturbations}
H_{\mu\nu}:=|g^{\mu\rho}|\,\delta h_{\rho\nu};\quad \partial_u H_{uu}:=\frac{4\,u\sqrt{f(u)}}{L^2}\delta h_{uu};\quad H_{\mu u}:= 2 \pi T\,|g^{\mu\rho}|\,\delta h_{\rho u},
\ee
where $g^{ab}$ is the inverse of the black brane metric (\ref{bb}). For definiteness, we aligned the momentum along the $x$-direction. The perturbations \eqref{eq.perturbations} are classified according to their transformation properties with respect to residual rotations $\mathcal{O}(2)$ in the plane transverse to their momentum, see, e.g., \cite{Kovtun:2005ev}. This gives rise to the scalar, vector and tensor channels, which, by construction, decouple from each other. Given that the tensor channel does not support the hydrodynamic (gapless) excitations, the corresponding modes are not going to contribute to the hydrodynamic effective action and we will neglect them. Hence, we are only going to consider the scalar and vector modes 
\ba\label{pert}
&&\text{Scalar (sound channel):} \quad H_{tt},\quad H_{xt},\quad H_{ii},\quad H_{aa},\quad H_{tu},\quad H_{xu},\quad H_{uu},\nn\\
&&\text{Vector (shear channel):}\quad H_{\alpha t},\quad H_{\alpha x},\quad H_{\alpha u},\quad\text{with}\quad \alpha=y,\,z.
\ea
For the future convenience, the formulas above utilized the following notation: 
\be
H_{ii}=H_{xx}+H_{yy}+H_{zz}\quad\text{and}\quad H_{aa}=H_{xx}-H_{yy}-H_{zz}.
\ee
Notice that our analysis here keeps arbitrary values of the lapse and shift variables, as opposed to the previous Section. This will allow us to be very explicit about the emergence of the Goldstone bosons on the gravity side.

\subsection{Shear channel}
The shear channel equations of motion are $E_{\alpha t}$ and $E_{\alpha x}$ ($ \alpha=y,\,z$) and take the form
\ba
&&H_{\alpha t}^{\prime\prime}-\frac{1}{u}H_{\alpha t}'-\t k^2\frac{1}{uf} H_{\alpha t}-\t k\,\t\omega\frac{1}{uf} H_{\alpha x}+i\t\omega \,H_{\alpha u}'-i\t\omega\frac{1}{u}H_{\alpha u}=0,\label{eqshear1}\\
&&H_{\alpha x}''\!-\!\frac{(1+u^2)}{uf}H_{\alpha x}'\!+\!\t \omega^2 \frac{1}{uf^2}H_{ \alpha x}\!+\!\t \omega\,\t k\frac{1}{uf^2}H_{\alpha t}\!-\!i\tilde{k}\,H_{\alpha u}'\!+\!i\tilde{k}\frac{(1+u^2)}{uf}H_{\alpha u}=0,\label{eqshear2}
\ea
where we defined the dimensionless frequency and momentum
\be
\tilde{\omega}=\frac{\omega}{2\pi T}\quad\text{and}\quad \tilde{k}=\frac{k}{2\pi T}.
\ee
The equations \eqref{eqshear1} and \eqref{eqshear2} need to be supplemented with the constraint $E_{\alpha u}$ 
\be\label{eqshear3}
\tilde{k}\, H_{\alpha x}^{\prime}+\frac{\tilde{\omega}}{f}H_{\alpha t}^{\prime}+i\frac{(\tilde{\omega}^2-f\tilde{k}^2)}{f}H_{\alpha u}=0.
\ee
Since we are interested in the low energy dynamics of the linearized perturbations, we will search for solutions in a perturbative derivative expansion. The fields $H_{\alpha u}$ naturally appear with a field theory derivative and they have to be retained at the same order as the other fields $H_{\alpha t}$ and $H_{\alpha x}$. We implement the gradient expansion by redefining $\omega\rightarrow \lambda \,\omega$, $k\rightarrow \lambda\, k$, rescaling the fields $H_{\alpha u}\rightarrow1/\lambda \,H_{\alpha u}$ and searching for solutions in a power series of the bookkeeping parameter $\lambda \ll 1$
\be\label{hydro}
H_{\mu\nu}=H_{\mu\nu}^{(0)}+\lambda\, H_{\mu\nu}^{(1)}+\lambda^2\, H^{(2)}_{\mu\nu}+\dots.
\ee
At this point, one usually fixes a gauge (typically, the radial gauge $H_{au}=0$) and solves the full set of equations (\ref{eqshear1}-\ref{eqshear3}) in the small-$\lambda$ expansion. In each transversal direction $\alpha$, the relevant equations are a set of two coupled second order ordinary differential equations and one first order equation. The total number of the integration constants per transverse direction is then three. They are usually fixed by setting two Dirichlet boundary conditions in the UV and imposing the ingoing boundary condition on the horizon. However, as in the previous Section, we want to solve here a double Dirichlet problem, namely we want to impose the Dirichlet boundary conditions not only in the UV but also on some IR brane. We are going then to solve \eqref{eqshear1} and \eqref{eqshear2} and leave the constraint \eqref{eqshear3} unsolved.
At leading order in the hydrodynamic expansion (\ref{hydro}) equations (\ref{eqshear1}-\ref{eqshear2}) are then
\ba
&&H_{\alpha t}^{(0)''}-\frac{1}{u}H_{\alpha t}^{(0)'}+i\tilde{\omega}\,  H_{\alpha u}'-i\tilde{\omega} \frac{1}{u}H_{\alpha u}=0,\\
&& H_{\alpha x}^{(0)''}-\frac{(1+u^2)}{uf}H_{\alpha x}^{(0)'}-i\tilde{k}\,H_{\alpha u}'+i\tilde{k}\frac{(1+u^2)}{uf}H_{\alpha u}=0. 
\ea
The solution with Dirichlet boundary conditions $H_{\mu\nu}^B$ in the UV ($u=0$) and Dirichlet boundary conditions $H_{\mu\nu}^{\delta}$ at some $u=u_{\delta}$ is not unique since it depends on the arbitrary gauge choice encoded in the fields $H_{\alpha u}$
\ba
H_{\alpha t}^{(0)}(u)&=&H_{\alpha t}^B-\frac{u^2}{u_{\delta}^2}\Delta H_{\alpha t}-i\tilde{\omega}\int_{0}^uH_{\alpha u}(w)\,dw,\label{sol1}\\
H_{\alpha x}^{(0)}(u)&=&H_{\alpha x}^B-\frac{\log f}{\log f_{\delta}}\Delta H_{ \alpha x}+i\tilde{k}\int_0^uH_{\alpha u}(w)\,dw\label{sol2}.
\ea
In the formula above, $f_{\delta}=f(u_{\delta})$ and we have also defined the following bulk diffeomorphisms invariant combinations 
\ba
\Delta H_{\alpha t}&=&H_{\alpha t}^B-H_{\alpha t}^{\delta}-i\tilde{\omega}\,\pi_{\alpha},\nn\\
\Delta H_{\alpha x}&=&H_{\alpha x}^B-H_{\alpha x}^{\delta}+i\tilde{k}\,\pi_{\alpha},
\ea
with $\pi_{\alpha}$ defined as a following Wilson line-like object
\be\label{Gold}
\pi_{\alpha}=\int_0^{u_{\delta}}H_{\alpha u}(u)\,du.
\ee

\subsubsection{The transverse Goldstones}

The Wilson line-like objects defined in (\ref{Gold}) are the (linearized) Goldstone bosons of certain spontaneously broken symmetries. In fact, one can easily see that the combinations (\ref{Gold}) are invariant under those bulk diffeomorphisms, which involve diagonal combinations of the diffeomorphisms on the two boundaries, and transform nontrivially otherwise.  
The gauge symmetry of reparametrizing the two Dirichlet boundary conditions  $\text{Diffs}_4\times\text{Diffs}_4$ is broken down to the diagonal combination $\text{diag}(\text{Diffs}_4)$ by the classical solution (\ref{sol1}-\ref{sol2}) and the Goldstones (\ref{Gold}) can be associated to the spontaneous breaking of  the global symmetry subgroup 
\be
\text{Poincar\'e}_4\times\text{Poincar\'e}_4\rightarrow \text{diag}(\text{Transl}_4+\text{Rot}_3).
\ee
In the formula above, the Lorentz group is broken completely as the two boundaries are characterized by different speed of light and only the diagonal combination of spacetime translations and rotations survive. 

If we work instead in a specific gauge, e.g. the radial gauge, the Goldstones (\ref{Gold}) arise as non-trivial boundary conditions to be imposed on the second boundary. For instance we can perform a bulk diffeomorphism $x^{a}\rightarrow x^{a}+\xi^{a}$ in order to transform the metric (\ref{bb}) with its perturbations $H_{ab}$ to a form where the new metric perturbation satisfy the condition $\t H_{aU}=0$ in the new bulk coordinates $y^a=(y^{\mu},U)$. Such diffeomorphism, in the lowest order in the derivative expansion, is
\be\label{diff}
\xi^{(0)}_{\alpha}(u)=\frac{1}{u}C_{\alpha}-\frac{1}{u}\int_0^u H_{\alpha u}(w)dw,
\ee
where  $C_{\alpha}=C_{\alpha}(\t \omega,\t k)$ does not depend on the radial direction and can be set to zero. The bulk metric perturbations change to
\ba
\tilde{H}_{\alpha t}^{(0)}(u)&=&H_{\alpha t}^{(0)}(u)+i\tilde{\omega}\int_0^u H_{\alpha u}(w)dw,\\
\tilde{H}_{\alpha x}^{(0)}(u)&=&H_{\alpha x}^{(0)}(u)-i\tilde{k}\int_0^u H_{\alpha u}(w)dw
\ea
and the boundary values transform accordingly
\ba
\tilde{H}_{t\alpha}^{\delta}&=&H_{t\alpha}^{\delta}+i\tilde{\omega}\,\pi_{\alpha},\\
\tilde{H}_{x\alpha}^{\delta}&=&H_{x\alpha}^{\delta}-i\tilde{k}\,\pi_{\alpha}.
\ea
Notice that  in the radial gauge the metric is of the form
\be \label{newform2}
ds^2 = dU^2 + 2 A_{\mu}(y^{\nu}) dy^{\mu}dU + g_{\mu\nu}(y^{\mu},U) dy^{\mu} dy^{\nu}
\ee
and the lines of constant $y^{\mu}$ are spatial geodesics with affine parameter $U$.
As described in the introduction, the Goldstone bosons (\ref{Gold}) correspond then to a map $x^{\mu}(y^{\mu},0)\rightarrow x^{\mu}(y^{\mu},u_{\delta})$ from the conformal boundary to the IR brane following suitable spatial geodesics.

\subsubsection{The transverse effective action}
Now that we have the solution of the double Dirichlet problem, we are ready to compute the 
 partially on-shell action between the IR and the UV brane. 
In order to make the variational problem well-defined we need to include the Gibbons-Hawking term on both of the boundaries (as in the previous Section) and a counterterm in the UV
\be\label{Stot}
S_{\delta}=S_{HE}|^{u_{\delta}}_{0}+S_{GH}|_{u_{\delta}}-S_{GH}|_{u=0}-S_{\text{ct}}|_{u=0},
\ee
where $S_{HE}$ is given in (\ref{HE}) and
\be
S_{GH}=\frac{1}{k_5^2}\int d^4x\sqrt{-\gamma}\,K;\quad \qquad 
S_{\text{ct}}=\frac{L}{2k_5^2 }\int d^4x\sqrt{-\gamma}\left(\frac{6}{L^2}+ ^{4\!\!\!}R\right),
\ee
In the formulas above, $\gamma$ is the determinant of the induced metric on the timelike hypersurface, $K$ is the trace of the extrinsic curvature tensor and $^{4\!\!}R$ is the Ricci scalar on the (3+1)-dimensional timelike hypersurface, which will only contribute in the second order of the derivative expansion. We will now set the action (\ref{Stot}) \emph{partially} on-shell by using the solutions (\ref{sol1}-\ref{sol2}). The background contribution takes the form 
\be\label{UVconst}
S_{\text{const}}=P_0V_4\left(3-\frac{6}{u^2_{\delta}}\right)\quad \text{with}\quad P_0=\frac{\pi^4 T^4 L^3}{8k_5^2},
\ee
where $V_4$ is the four-dimensional volume term and $P_0$ is the thermodynamic pressure. The contribution of the perturbation is given by
\ba\label{UVshear}
S_{T}&=&-P_0V_2\int\frac{dk\,d\omega}{(2\pi)^2}\sum_{\alpha}\Bigg(\frac{3}{2}(H_{\alpha t}^{B})^2+\frac{1}{2}(H_{\alpha x}^B)^2+\nn\\
&&+\frac{3}{u_{\delta}^2}(H_{\alpha t}^{\delta})^2-\frac{(2+f_{\delta})}{u_{\delta}^2}(H_{\alpha x}^{\delta})^2-\frac{2}{u_{\delta}^2}(\Delta H_{\alpha t})^2-\frac{2}{\log f_{\delta}}(\Delta H_{\alpha x})^2\Bigg),
\ea
where $V_2$ is the two-dimensional transverse volume term and we have omitted the arguments of the fields for which we use the convention
\be\label{conv}
A\,B=\frac{1}{2}\left(A(\t\omega,\t k)B(-\t\omega,-\t k)+A(-\t\omega,-\t k)B(\t\omega,\t k)\right).
\ee
The equations of motion for the Goldstone fields, as derived from the effective action (\ref{UVshear}), correspond to the constraint equations (\ref{eqshear3}) and represent conservation of the energy-momentum tensor in the dual field theory. Imposing vanishing double Dirichlet boundary conditions, the effective action will depend only on the Goldstone degrees of freedom
\ba
S_{\pi^T}&=&P_0 V_2\int\frac{dk\,d\omega}{(2\pi)^2}\sum_{\alpha}2\left(\frac{\t\omega^2}{u_{\delta}^2}+\frac{\t k^2}{\log f_{\delta}}\right)\pi_{\alpha}^2.
\ea
The linear dispersion relation is immediately derived 
\be
\t\omega_T=\pm \,c_T\,\t k,\quad \text{with}\quad c_T=\frac{u_{\delta}}{\sqrt{-\log f_{\delta}}},
\ee 
and depends on the position $u_{\delta}$ of the IR brane suggesting that on  
a finite cutoff $u_{\delta}$ the volume preserving diffeomorphisms is broken. This is very much in line with the analysis presented in the previous Section.

In the near horizon limit $u_{\delta}\rightarrow 1$ the background on-shell action
\be
S_{\text{const}}|_{\cal H}= - 3\, P_0\,V_4,
\ee
represents the energy density times the four-volume of a holographic conformal fluid.
The transverse effective action
\ba
S_{T}|_{\cal H}&=&P_0V_2\int\frac{dk\,d\omega}{(2\pi)^2}\sum_{\alpha}\Bigg(\frac{1}{2}(H_{\alpha t}^{B})^2-\frac{1}{2}(H_{\alpha x}^B)^2-4\, H_{\alpha t}^B\,H_{\alpha t}^{\delta}+2\,(H_{\alpha x}^{\delta})^2-(H_{\alpha t}^{\delta})^2+\nn\\
&&\qquad \qquad \qquad +2\,i\t \omega \,\big((H_{\alpha t }^B-H_{\alpha t }^{\delta})\,\pi_{\alpha}-\pi_{\alpha}(H_{\alpha t }^B-H_{\alpha t }^{\delta})\big)+2\, \t \omega^2\pi_{\alpha}^2\Bigg)
\ea
turns out to be equivalent to the Fourier transform of the transverse sector in Eq.~(\ref{effactionfull}) derived in Section \ref{TimeGold} when the boundary metric expansion is included. In order to demonstrate it, one needs to redefine $\pi_{\alpha}\rightarrow -\pi_{\alpha}$, impose the conformal fluid equation of state $F(s)=-s^{4/3}$ and set $s_0$ to $s_0^{4/3} \equiv 3P_0$. Furthermore, one also needs to add the contribution $+3\,(H_{\alpha t}^{\delta})^2$ coming from the difference between the near horizon form of the metric (\ref{bb}) with linear perturbations and the Galilean form of the horizon metric (\ref{IRmetric}), where in the $tt$-component the first nontrivial term is second order in an amplitude expansion. Notice also that in the near horizon limit 
 the transverse velocity $c_T\rightarrow 0$ and the trivial shear waves dispersion relation (\ref{sheardisp}) is recovered. 

At higher orders of the hydrodynamic expansion, the effective action contains divergent terms as the stretched horizon approaches the position of the event horizon. However, the resulting dispersion relation for the Goldstones is trivial  
\be
\t\omega_{T}=\mathcal{O}(1-u_{\delta})+\mathcal{O}(\tilde{k}^4),
\ee
and does not retain any of the aforementioned undesired features
if one is careful in taking the near horizon limit at each order of the hydrodynamic expansion. The reason for it is simply that the two limits do not commute.

To recap, we demonstrated here that up to the second order of hydrodynamic gradient expansion, the shear mode does not propagate provided one ignores the dissipative effects. This hints towards the volume-preserving diffeomorphisms invariance being the symmetry of the effective action for holographic fluids at least up to the second order of the gradient expansion.

\subsection{Sound channel}

All the manipulations of the previous Section can be repeated pretty much straightforwardly also for the sound channel perturbations. The dynamical Einstein's equations ($E_{\mu\nu} = 0$) in the leading order of the gradient expansion take the form
\ba
&&H_{xt}^{(0)''}-\frac{1}{u}H_{xt}^{(0)'}-i\t k\,f\,H_{tu}'+i\t k\frac{(1+3u^2)}{u}H_{tu}+i\t\omega\, H_{xu}'-i\t\omega\frac{1}{u} H_{xu}=0,\label{eqsound1}\\
&&H_{aa}^{(0)''}-\frac{(1+u^2)}{uf}H_{aa}^{(0)'}-2\,i\t k\,H_{xu}'+2\,i\t k\frac{(1+u^2)}{uf}H_{xu}=0,\\
&&H_{ii}^{(0)''}-\frac{1}{uf}H_{ii}^{(0)'}-2\,i\t k\,H_{xu}'+2\,i\t k\frac{1}{uf}H_{xu}+\frac{3}{2}\sqrt{f}H_{uu}''-\frac{3}{2}\frac{(1+2u^2)}{u\sqrt{f}}H_{uu}'=0,\\
&&H_{tt}^{(0)''}-\frac{(1+2u^2)}{uf}H_{tt}^{(0)'}-\frac{2}{3}H_{ii}^{(0)''}+\frac{2}{3}\frac{(1+u^2)}{uf}H_{ii}^{(0)'}+2\,i\t \omega \,H_{tu}'-2\,i\t\omega\,\frac{(1+2 u^2)}{uf}H_{tu}+\nn\\
&&\quad +\frac{4}{3}i\t k\,H_{xu}^\prime -i\t k\frac{4(1+u^2)}{3uf}H_{xu} -\frac{(3-u^2)}{2\sqrt{f}}H_{uu}''+(1+u^2)\frac{(3-2u^2)}{2uf^{3/2}}H_{uu}'=0\label{eqsound4},
\ea
and the constraint equations $E_{\mu u}$  read
\ba
&&i\t k \!\left(H_{xt}^{(0)\prime}\!+\!\frac{2u}{f}H_{xt}^{(0)}\right)\!+\!i\t\omega\left( H_{ii}^{(0)\prime}+\frac{u}{f}H_{ii}^{(0)}+\frac{3}{2}\sqrt{f} H_{uu}^{\prime}\right)\!+\!\t k^2fH_{tu}\!+\!\t k\t\omega \,H_{xu}=0,\quad\quad\label{eqsound5}\\
&&i\t k\!\left(\!\!H_{tt}^{(0)\prime}\!\!-\!\frac{u}{f}H_{tt}^{(0)}\!+\!\frac{2}{3}(H_{aa}^{(0)\prime}\!\!-\!\!H_{ii}^{(0)\prime})\!-\!\frac{(3-u^2)}{2\sqrt{f}}H_{uu}^\prime\!\!\right)\!\!+\!\frac{i\t \omega H_{xt}^\prime}{f}\!-\!\t k\t\omega H_{tu}\!\!-\!\frac{ \t\omega^2 H_{xu}}{f}\!=\!0,\quad\quad\label{eqsound6}\\
&&H_{tt}^{(0)\prime} -\frac{(3-u^2)}{3f}H_{ii}^{(0)\prime}+2\,i\t \omega \,H_{tu}+i\t k\frac{2(3-u^2)}{3f}H_{xu}-\frac{2}{\sqrt{f}}H_{uu}^\prime=0.\quad\quad\label{eqsound7}
\ea
The solutions to (\ref{eqsound1}-\ref{eqsound4}) with double Dirichlet boundary conditions depend on, basically freely-specifiable, values of $H_{tu}, H_{xu}$ and $H_{uu}$
\ba
H_{xt}^{(0)}(u)&=&H_{xt}^B-\frac{u^2}{u_{\delta}^2}\Delta H_{tx}+i\t k\,f\int_{0}^uH_{tu}(w)dw-i\t\omega\int_0^uH_{xu}(w)dw,\nn\\
H_{ii}^{(0)}(u)&=&H_{ii}^B-\frac{1-\sqrt{f}}{1-\sqrt{f_{\delta}}}\Delta H_{ii}+2\,i\t k\int_0^uH_{tu}(w)dw-\frac{3}{2}f\,H_{uu}(u),\nn\\
H_{aa}^{(0)}(u)&=&H_{aa}^B-\frac{\log f}{\log f_{\delta}}\Delta H_{aa}+2\,i\t k\int_0^uH_{xu}(w)dw,\nn\\
H_{tt}^{(0)}(u)&=&H_{tt}^B-\frac{\sqrt{f_{\delta}}(1-\sqrt{f})}{\sqrt{f}(1-\sqrt{f_{\delta}})}\Delta H_{tt}+\frac{1}{3}\frac{(1-\sqrt{f})(\sqrt{f_{\delta}}-\sqrt{f})}{\sqrt{f}(1-\sqrt{f_{\delta}})}\Delta H_{ii}+\nn\\
&&-2\,i\t\omega\int_0^uH_{tu}(w)dw+\frac{1+u^2}{2\sqrt{f}}H_{uu}(u),
\ea
where we have defined the following bulk diffeomorphisms invariant combinations
\ba
\Delta H_{xt}&=&H_{xt}^B-H_{xt}^{\delta}+i\t k\,f_{\delta}\,\pi_t-i\t \omega \,\pi_x,\nn\\
\Delta H_{ii}&=&H_{ii}^B-H_{ii}^{\delta}+2\,i\t k\,\pi_x-\frac{3}{2}\sqrt{f_{\delta}}\,H_{uu}(u_{\delta}),\nn\\
\Delta H_{aa}&=&H_{aa}^B-H_{aa}^{\delta}+2\,i\t k\,\pi_x,\nn\\
\Delta H_{tt}&=&H_{tt}^B-H_{tt}^{\delta}-2\,i\t\omega\, \pi_t+\frac{1+u_{\delta}^2}{2\sqrt{f_{\delta}}}H_{uu}(u_{\delta}).
\ea
In complete analogy with the previous Section, we also defined the following (linearised) Goldstones
\be
\pi_{t}=\int_0^{u_{\delta}}H_{tu}(u)du\quad \mathrm{and} \quad\pi_{x}=\int_0^{u_{\delta}}H_{xu}(u)du.
\ee
Notice that the contribution $H_{uu}$ appears here with no derivatives. This metric component is in fact non-dynamical and it is associated to the parametrization of the position of the IR brane $u_{\delta}$.

\subsubsection{The longitudinal effective action}
The on-shell action (\ref{Stot}) up to second order in an amplitude expansion in the sound channel with vanishing double Dirichlet boundary conditions is
\ba\label{longgold}
S_{\pi^L}&=&P_0\,V_2\int\frac{dk \,d\omega}{(2\pi)^2}\Bigg( \frac{f_{\delta}}{u_{\delta}^2}\left(2\,\t k^2-3\,\t \omega^2\right)\pi_t^2-2\,\t \omega\,\t k\frac{f_{\delta}}{u_{\delta}^2}\pi_t \, \pi_x+\nonumber\\
&&+\,\frac{\left(8\,\t k^2u_{\delta}^2-\left(\t k^2(1+u_{\delta}^2)-6\,\t\omega^2\right)\log f_{\delta}\right)}{3\,u_{\delta}^2\,\log f_{\delta}}\pi_x^2\Bigg),
\ea
where we followed the same convention as in Eq.~(\ref{conv}). Notice that the contribution of $H_{uu}$ was integrated out. If the IR brane is kept at an arbitrary radial position $u_{\delta}$, both Goldstones $\pi_t$ and $\pi_x$ are dynamical with coupled equations of motion 
\ba\label{actiongold}
&&\frac{f_{\delta}}{u_{\delta}^2}\left(2\,\t k^2-3\,\t \omega^2\right)\pi_t-\t \omega\,\t k\frac{f_{\delta}}{u_{\delta}^2}\pi_x=0,\\
&&\t \omega\,\t k\frac{f_{\delta}}{u_{\delta}^2}\pi_t+\frac{\left(8\,\t k^2u_{\delta}^2-\left(\t k^2(1+u_{\delta}^2)-6\,\t\omega^2\right)\log f_{\delta}\right)}{3\,u_{\delta}^2\,\log f_{\delta}}\pi_x=0.
\ea
As previously, these equations correspond to the constraint equations, here Eq.~\eqref{eqsound5} and \eqref{eqsound6}, and hence follow from the conservation of the dual energy-momentum tensor. We can now solve Eq.~\eqref{actiongold} for the dispersion relations. We obtain two modes, which decouple in the vicinity of the event horizon and correspond \emph{then} to the independent oscillations of $\pi_{t}$ and $\pi_{x}$:
\ba
\pi_t:\qquad \t\omega&=&\pm\,\sqrt{\frac{2}{3}}\,\t k+\mathcal{O}(\t k^3)\label{pit}\\
\pi_x:\qquad\t\omega_L&=&\pm\,\frac{1}{\sqrt{3}}\t k+\mathcal{O}(1-u_{\delta})+\mathcal{O}(\t k^3).
\ea
Notice that the longitudinal Goldstone $\pi_x$ has the standard dispersion relation for sound waves, see Eq.~\eqref{eq.speedofsound}. The other mode, discussed previously in \cite{deBoer:2014xja}, is not present in relativistic hydrodynamics. In fact, it is easily seen from the effective action point of view (\ref{longgold}) that in the near-horizon limit $u_{\delta}\rightarrow 1$ all $\pi_t$ contributions vanish and only the longitudinal mode $\pi_x$ survives. Hence, although the dispersion relation (\ref{pit}) is finite on the horizon, it is associated with unphysical mode and has to be discarded\footnote{Another way to see this is by looking at the residue of the resulting two-point function for the dual energy-momentum tensor. The residue related to the pole (\ref{pit}) vanishes in the near-horizon limit and as a result the corresponding mode disappears. We thank Dam T. Son for pointing this out.}.

Going to higher order in the hydrodynamic gradient expansion at the level of the effective action is technically quite demanding. Despite that, it is still possible to solve the double Dirichlet problem and investigate the constraint equations, which we did up to the second order in a derivative expansion. Proceeding in this way, we derived the correction to the dispersion relation for the longitudinal sector 
\be\label{dispsound1}
\t \omega_L=\pm\,\frac{1}{\sqrt{3}}\tilde{k}\pm\left(\frac{2}{3\sqrt{3}}+\frac{\log (1-u_{\delta})}{18\sqrt{3}}-\frac{5\log 2}{18\sqrt{3}} \right)\tilde{k}^3+\mathcal{O}(1-u_{\delta})+\mathcal{O}(\t k^4).
\ee
Notice that although such dispersion relation is purely real and, hence, dissipationless,
it diverges in the near-horizon limit. We expect the corresponding divergence to appear in the effective action, although we did not check this explicitly. It is hard to interpret this divergence univocally. Perhaps the most straightforward interpretation is that beyond the leading order in the gradient expansion keeping the vanishing Dirichlet boundary conditions on the event horizon is unphysical. A more speculative interpretation is that at the level of the holographic correspondence it is simply not possible to split the fluid into the dissipative and dissipationless part. Leaving this for future investigations, we finish this Section by pointing out that the divergent contributions to \eqref{dispsound1} are intrinsically associated with the $\omega$-dependence. Hence, it is natural to expect that in the Euclidean setting in thermal
equilibrium such divergences are absent and that the action functional for
fluids (also beyond the leading order in the gradient expansion) is well-defined.

\section{Coupling to an IR sector\label{sec5}}

So far we dealt only with the part of the spacetime between some IR and UV branes, ultimately sending one of the cutoffs to the UV boundary and trying to send the other to the event horizon. However we never included the very important property of the horizon being a surface of no return, i.e. we never included the dynamical contributions of the part of the spacetime between the horizon and the IR brane. Having an intermediate cutoff $u_{\delta}$  naturally splits the spacetime into a UV and IR sector and, as a consequence, the bulk action also splits into two parts
\be
S=S^{IR}+S^{UV}=\frac{1}{2k_5^2}\int^{1}_{u_{\delta}} du\,d^4x\,\sqrt{-g}\, (R-2\Lambda)+\frac{1}{2k_5^2}\int^{u_{\delta}}_{0} du\,d^4x\,\sqrt{-g}\, (R-2\Lambda).
\ee
The (partially) on-shell UV part of the action computed in a derivative expansion is what acquires the interpretation of the effective action for dissipationless hydrodynamical excitations, at least at the leading order. In order to couple such action to the IR sector, one needs to integrate out the IR fields on the finite cutoff  $H_{\mu\nu}^{\delta}$
\be\label{IRdata}
\frac{\delta S}{\delta \,H_{\mu\nu}^{\delta}}=\frac{\delta S^{IR}}{\delta \,H_{\mu\nu}^{\delta}}+\frac{\delta S^{UV}}{\delta \,H_{\mu\nu}^{\delta}}=0
\ee
and setting there the Dirichlet boundary conditions has to be understood as a useful intermediate step \cite{Faulkner:2010jy,Heemskerk:2010hk,Nickel:2010pr}.

In the remaining part of this Section we are going to focus on two different ways to couple the UV sector to the IR. First, we will use a membrane paradigm approximation and derive the usual damped dispersion relation for the sound waves, without any trace of the divergence discussed in the previous Section (see~Eq.~\eqref{dispsound1}). Secondly, we will focus on static configuration and employ the coupling to a regular Rindler-type dynamical sector, providing the first derivation of the hydrodynamic partition function from holography.

\subsection{Coupling to the membrane paradigm: dissipation\label{sec5.mp}}

In order to recover dissipation, it is clearly necessary to include the horizon contribution and ultimately recovering the ingoing boundary condition. Following Ref.~\cite{Nickel:2010pr,Faulkner:2010jy}, instead of retaining the full dynamical IR sector and dealing with the Schwinger-Keldish formalism, we are going to use the membrane paradigm approximation. To achieve this, we will impose a convenient boundary condition on a finite cutoff $u_{\delta}$
\be\label{membrane}
2(1-u)\frac{Z'(u)}{i \t\omega \,Z(u)}\Bigg|_{u_{\delta}}=\sigma\quad \text{with}\quad \sigma=1,
\ee
where $Z$ is the relevant gauge invariant gravitational perturbation. Since we are concentrating on the hydrodynamical excitations, the membrane paradigm is a good approximation. See our previous paper \cite{deBoer:2014xja} for an extensive discussion on this point.

The nonlocal gauge-invariant combinations are
\ba\label{gaugeinv}
Z_{\alpha}^T&=& \tilde{k}\, H_{t\alpha}+\t \omega \,H_{x\alpha},\nn\\
Z^L&=&2\,\t k^2f\,H_{tt}+4 \,\t\omega\, \t\, kH_{xt}+2\,\t\omega^2H_{xx}+H_{aa}\left(\t k^2(1+u^2)-
\t \omega^2\right),
\ea 
respectively in the shear and sound channels. Keeping now the IR Dirichlet boundary conditions $H_{\mu\nu}^{\delta}$ non-vanishing, solving the constraint equations with respect to the Goldstones and using Eq. (\ref{membrane}) gives the dispersion relations for the shear and sound modes
\ba
\t\omega_T&=&-\frac{i}{2}\sigma \,\t k^2-\frac{i}{8}\sigma\left(2+ (1-\sigma^2)\log(1-u_{\delta})-(1+\sigma^2)\log 2\right)\t k^4+\mathcal{O}(1-u_{\delta}),\\
\t\omega_L&=&\pm\,\sqrt{\frac{1}{3}} \,\t k-\frac{i}{3}\sigma\, \tilde{k}^2+\nonumber\\
&&
\pm\left(\frac{1}{2\sqrt{3}}-\frac{\log2}{3\sqrt{3}}+(1-\sigma^2)\frac{(1+\log2+\log(1-u_{\delta}))}{6\sqrt{3}}\right)\t k^3+\mathcal{O}(1-u_{\delta}),
\ea
as a function of the membrane coupling $\sigma$.

Notice that decoupling the membrane by setting the membrane coupling $\sigma=0$ gives a dissipationless dispersion relation which, however, does not coincide with Eq.~(\ref{dispsound1}). There is a simple explanation to this. From Eq.~(\ref{membrane}) it follows that imposing $\sigma=0$ corresponds to setting Neumann rather than Dirichlet boundary condition on the IR brane. This result demonstrates that also for a different set of boundary conditions we do get the divergent terms in the dispersion relation for sound waves. Several boundary conditions could in principle give different dissipationless effective actions and dispersion relations, as long as we make sure there is no net flux through the IR brane. The divergent logarithmic term is removed when the ingoing boundary conditions ($\sigma=1$) are imposed, reproducing the correct dispersion relation found earlier in the literature, see, e.g., \cite{Baier:2007ix}.
This complements our discussion from the previous Section on the division of holographic fluids into dissipative and non-dissipative contribution.

\subsection{Coupling to an Euclidean IR sector: the equilibrium partition function\label{sec5.epf}}

The equilibrium partition function of \cite{Banerjee:2012iz,Jensen:2012jh} can be computed holographically by evaluating the on-shell action on solutions to Einstein's equations in the Euclidean signature with arbitrary boundary metrics and a regular boundary condition at the tip of the cigar.
Our setup can be viewed as an intermediate step to obtain the same result. In fact, this can be achieved by coupling the effective action in the static limit $\omega\rightarrow 0$ to an Euclidean IR sector which takes care of the near horizon (regular) region of the spacetime.
Notice that in the conventional derivation of the thermodynamic partition function from gravity there is no Gibbons-Hawking term in the IR, while in the effective action formalism we had to retain such term in (\ref{Stot}) since it was non-vanishing in the near horizon limit.
It is then natural to expect that the IR sector is proportional to such a contribution and we will show in the following that this is, in fact, the case.

Consider a regular cigar-shaped geometry, which near the horizon of a black hole  looks like  the tip of the cigar times the horizon geometry
\be \label{ir2}
ds^2 = \frac{\beta_{IR}^2 G_{tt}}{(2\pi)^2 r_0^2}\left((dr^2 + \frac{(2\pi )^2}{\beta_{IR}^2} r^2 (d\phi^t)^2\right)
+ G_{ij}(d\phi^i-v^i d\phi^t) (d\phi^j-v^j d\phi^t),
\ee
where we assumed Euclidean time has periodicity $\beta_{IR}$. Setting $r=r_0$ we recover the Euclidean metric on the IR brane
\be \label{ir1}
ds^2 = G_{MN} d\phi^M d\phi^N= G_{tt} d\phi^t d\phi^t + 
G_{ij} (d\phi^i-v^i d\phi^t) (d\phi^j-v^j d\phi^t).
\ee
The geometry (\ref{ir2}) does not solve the Einstein's equations, but since we are working at the leading order in
derivatives this does not matter. Moreover, we will assume that $r_0$ is very small with $G_{tt}\sim r_0^2$.
We denoted the inverse temperature by $\beta_{IR}$ to emphasize that this is the temperature as seen by the IR metric,
which is not necessarily the same as the temperature defined by the UV metric.

The on-shell value of the Euclidean action that covers the near horizon region
$0\leq r\leq r_0$ contains in principle two contributions
\be
S^{IR}=S_{HE}\Big|^{0}_{r_0}+S_{GH}\Big|_{r_0}.
\ee
The bulk Einstein-Hilbert action scales as $ S_{HE}\sim \mathcal{O}(r_0)$ since the integration domain shrinks to zero.  
The Gibbons-Hawking term turns out to be independent of $r_0$ and equal to
\be \label{gh}
S_{GH}= \int d^d \phi^M \,\frac{\sqrt{\det G_{ij}}}{\beta_{IR}}.
\ee
To proceed, we make a change of coordinates $(\phi^i-v^i \phi^t )\rightarrow \phi^i$ which we can always undo
later. Since we are working at the lowest order in derivatives we can assume the $v^i$ to be constant, and the
change of coordinates therefore removes the $d\phi^t d\phi^i$ cross terms from the metric.
We can then rewrite (\ref{gh}) as
\be \label{gh2}
S_{GH}= \int d^d \phi^M\, \frac{\sqrt{\det G_{MN}}}{\beta_{IR}\sqrt{G_{tt}}} = \int d^d x \frac{\sqrt{\det h}}{\beta_{IR}\sqrt{G_{tt}}},
\ee
where $h$ is defined in (\ref{eq.defh}).
If we denote 
$$\Sigma^{MN}=\frac{\partial\phi^M}{\partial x^{\mu}} 
\frac{\partial\phi^N}{\partial x^{\nu}} g^{\mu\nu},
$$
then we can use the fact that $G_{MN}$ is block diagonal to deduce the following identity
\be
\det(\Sigma^{MN} G_{NK}) = \frac{\det(\Sigma^{ij} G_{jk}) \sqrt{G_{tt}} }{(\Sigma^{-1})_{tt}}.
\ee
If we insert this identity in (\ref{gh2}) we obtain
\be \label{gh3}
S_{GH}= 
\int d^d x \sqrt{g} \det\left( \frac{\partial\phi^i}{\partial x^{\mu}} 
\frac{\partial\phi^j}{\partial x^{\nu}} g^{\mu\nu} G_{jk}\right)^{1/2} \frac{1}{\beta_{IR}\, {\sigma}},
\ee
where
\be \sigma^2 = (\Sigma^{-1})_{tt}
= 
g_{\mu\nu} \frac{\partial x^{\mu}}{\partial\phi^t} \frac{\partial x^{\nu}}{\partial\phi^t}.
\ee
The quantity $\sigma$ has a simple interpretation: it is the norm of the vector field $\frac{\partial}{\partial \phi^t}$
pulled back to the UV boundary. Therefore, $\beta_{IR}\, {\sigma}$ is the proper length of the Euclidean time circle as
perceived on the UV boundary. We will therefore take
\be
\beta_{UV} = {\sigma}\, \beta_{IR} 
\ee
as our definition of the inverse UV temperature.
With this definition, we now see that  
\be \label{gh4}
S_{GH}= 
\int d^d x \sqrt{g} \det\left( \frac{\partial\phi^i}{\partial x^{\mu}} 
\frac{\partial\phi^j}{\partial x^{\nu}} g^{\mu\nu} G_{jk}\right)^{1/2} \frac{1}{\beta_{UV}} = 
\int d^d x \sqrt{g} \frac{s}{\beta_{UV}},
\ee
where in the last line we 
reinstate the $v^i$-dependence by undoing the coordinate transformation $(\phi^i-v^i \phi^t) \rightarrow \phi^i$ to recover
 precisely the entropy density as defined in Eq.~(\ref{entropynew}).
Hence, to summarize, we have just shown that the relevant  contribution of the IR sector in the near horizon limit is given by the Gibbons-Hawking term (\ref{gh4}). Most importantly, it is of the form $S^{IR}\sim T s\, V_d$, where $s$ is the entropy density, $T$ is the temperature of the fluid and $V_d$ is the spacetime volume.

Now, as promised, we couple the IR action (\ref{gh4}) to the UV effective action
derived in Section \ref{sec4} in the static limit $\omega\rightarrow 0$. The coupling is realized by integrating out IR data as required in (\ref{IRdata}), which effectively sets the Goldstones on-shell. Notice also that since $S^{UV}\sim - \epsilon\,V_d$ where $\epsilon$ is the energy density, we are actually performing a Legendre transform of the energy density with respect 
 to the entropy density  which gives the pressure $P=Ts-\epsilon$ as a function of T. With arbitrary background metric configurations and using the notation of Section \ref{sec4} the final result is
\ba\label{pfminwalla}
S&=&P_0 V_4+P_0 V_3\int d x\left(\frac{3}{2}H_{tt}^B+\frac{1}{2}H_{ii}^B\right)+\nn\\
&&+P_0V_3\int dx\Bigg(\frac{15}{8}(H_{tt}^B)^2+\frac{1}{2}(H_{xt}^B)^2-\frac{1}{8}(H_{xx}^B)^2+\frac{3}{4}H_{tt}^B H_{ii}^B+\frac{1}{2}H_{xx}^B H_{yy}^B+\nn\\
&&+\frac{1}{2}\sum_{\alpha}\left((H_{\alpha t}^B)^2-(H_{\alpha x}^B)^2\right)
\Bigg).
\ea
As expected such expression corresponds to the equilibrium partition function 
\be\label{PF}
\beta F=-\ln Z=-\int d^dx \sqrt{-g}\, P(T),
\ee
where $Z$ is the partition function of the system, $\beta=1/T$ is the inverse of the temperature $T$, $P(T)$ is the pressure as a function of the temperature $T$ and $g$ is the background fluid metric, which is assumed to have a timelike Killing vector such that it is time-independent. 
In the case of a conformal fluid, $P(T)= c \, T^{d} $.  Expressions (\ref{pfminwalla}) and  (\ref{PF}) match when the general background metric in 3+1 dimensions is a linearized perturbation around Minkowski metric and only shear and sound channels are taken into account
\be
ds^2=-(1- H_{tt}^B(x))dt^2+2\, H_{it}^B(x) dx^i dt+(\delta_{ij}+ H_{ij}^B(x))dx^idx^j.
\ee
The temperature is $T=T_0/\sqrt{1- H_{tt}^B}$ and the constant $c$ is fixed to match the equilibrium pressure: $P_0=c\,T_0^4$.

\section{The entropy current as a Noether current\label{sec6}}

In this Section we want to explore the
 the role that  the conserved entropy current
$J^{\mu} = s u^{\mu}$ plays in our setup. It turns out that the entropy current is related to a symmetry as in \cite{Haehl:2014zda,Haehl:2015pja}. To describe this symmetry, we put $v^i=0$ for simplicity and first define an IR stress tensor
\be \label{deftir}
T_{IR}^{MN} =- \frac{2}{\sqrt{-G}} \frac{\delta S}{\delta G_{MN}} \det\left( \frac{\partial x^{\mu}}{\partial \phi^M}
\right).
\ee
The extra determinant has been put in because we want the IR stress tensor to be defined with respect to the measure
$d^d \phi^M$ and not with respect to $d^d x$. Just as we do in fluids in Landau frame, we can look for a unit timelike
eigenvector $u_{IR}^M$ of $T_{IR}$ which obeys
\be \label{eigenv}
T_{IR}^{MN} (u_{IR})_N = - \rho_{IR}\, u_{IR}^M.
\ee
We can in principle find the eigenvalue $\rho_{IR}$ using the explicit form of the near-horizon metric (\ref{ir2}), 
and using the fact that the derivative of the effective action with respect to a boundary metric is proportional 
to the conjugate momentum,
or radial derivative, of that metric;  
however, we do not need the explicit form of $\rho_{IR}$ in our analysis below. We now claim that whenever
\be \label{symm}
\phi^M \rightarrow \phi^M + \frac{u_{IR}^M}{\rho_{IR}}
\ee
is a symmetry of the action, the corresponding conserved current is precisely the entropy current.

To show this, we first observe that for our action, which was of the type
\be \label{aux91}
S= \int d^d x \sqrt{-g} F[g_{\mu\nu},h_{\mu\nu}],
\ee
with $h$ given in (\ref{eq.defh}). The covariantly conserved Noether current for a transformation of the type (\ref{symm}) is
\be
j^{\mu} = 2 \frac{\delta F}{\delta h_{\mu\nu}} G_{MN} \frac{\partial \phi^M}{\partial x^{\nu}} \frac{u_{IR}^N}{\rho_{IR}} =
2 \frac{\delta F}{\delta G_{MN}} \frac{\partial x^{\mu}}{\partial \phi^M} G_{NK} \frac{u_{IR}^K}{\rho_{IR}}.
\ee
Using the definition of the IR stress tensor in (\ref{deftir}) and the eigenvalue 
equation (\ref{eigenv}) this becomes
\be
j^{\mu} =  \frac{\sqrt{-G}}{\sqrt{-g}} \det\left( \frac{\partial \phi^M}{\partial x^{\mu}} \right)  
\frac{\partial x^{\mu}}{\partial \phi^N} u_{IR}^N .
\ee

We now perform a near-horizon limit specializing to the case where $u_{IR}^M= \delta^{Mt}/\sqrt{-G_{tt}}$ is a vector purely in the $\phi^t$-direction.
The conserved current is then
\be
j^{\mu} =  \frac{\sqrt{-G}}{\sqrt{-g}} \det\left( \frac{\partial \phi^M}{\partial x^{\mu}} \right)  
\frac{\sigma}{\sqrt{-G_{tt}}} u_{UV}^{\mu},
\ee
where we introduced the unit vector
\be
u_{UV}^{\mu} = \frac{1}{\sigma}  \frac{\partial x^{\mu}}{\partial \phi^t},
\ee
which can be thought of as the suitably normalized pull back of the IR vector $u_{IR}$.

If we look back at our analysis of the IR effective action, in particular at (\ref{gh2}) and (\ref{gh4}), we 
see that one way to write the entropy density $s$ is as
\be 
s = \frac{\sqrt{-G}}{\sqrt{-g}} \det\left( \frac{\partial \phi^M}{\partial x^{\mu}} \right)  
\frac{\sigma}{\sqrt{-G_{tt}}}
\ee
and therefore 
\be \label{currfin}
j^{\mu} =  s\, u_{UV}^{\mu} ,
\ee
which is indeed the same as the entropy current.

Strictly speaking, we are not quite done at this point, because we should also show that $u_{UV}^{\mu}$ is the fluid
velocity. This can be demonstrated as follows. Because the function $F$ in (\ref{aux91}) must be a scalar and
does not involve derivatives, it must be a function of traces of products of $h_{\mu\nu}$ and $g^{\mu\nu}$. This implies
in particular that it obeys the equation
\be
\frac{\partial F}{\partial g_{\mu\rho}} g_{\rho\nu} + 
\frac{\partial F}{\partial h_{\mu\rho}} h_{\rho\nu} = 0.
\ee
It is not difficult to see that this equation implies that if $u_{IR}^M$ is an eigenvector of $T^{MN}_{IR}$, then
\be
u_{UV}^{\mu} = \frac{\partial x^{\mu}}{\partial \phi^M} u_{IR}^M
\ee
is automatically an eigenvector of $T^{\mu\nu}_{UV}$, the stress tensor obtained by varying the action with respect
to $g_{\mu\nu}$. Therefore, the vector $u_{UV}^{\mu}$ appearing in (\ref{currfin}) is automatically an eigenvector
of the UV stress tensor and therefore precisely equal to the fluid velocity in Landau frame. 

To summarize we have shown that the Noether current associated to the symmetry (\ref{symm}), with $u_{IR}^M$ the unit eigenvector
of the IR stress tensor $T_{IR}^{MN}$ defined in (\ref{deftir}) with eigenvalue $\rho_{IR}$ as defined in (\ref{eigenv})
is precisely the entropy current of the system. 

It is interesting that our system appears to have two temperatures, two stress tensors, and two fluid velocities, defined
with respect the IR and UV boundary respectively as in \cite{Haehl:2014zda,Haehl:2015pja}. This is perhaps an automatic consequence of our setup where the two
boundaries appeared on equal footing. In the limit where the IR boundary becomes very close to the horizon of a black hole, 
the IR fluid physics becomes quite simple, as it is governed by the universal near-horizon Rindler region. These simple
properties are then propagated to the UV boundary with the help of the Goldstone bosons. In particular, the entropy, which
in the near-horizon region is very simple and proportional to the area of the horizon, becomes somewhat more involved
when described in terms of the UV variables\footnote{See \cite{Bhattacharyya:2008xc,Booth:2010kr} for earlier constructions of hydrodynamic entropy currents from gravity. These results were directly motivated by area theorems.}. We have also explained how the entropy current can be associated to a symmetry
which is purely based on the IR variables. This symmetry corresponds to some type of invariance of the IR dynamics as
one flows along with the IR fluid velocity, with a suitable normalization. It would clearly be very interesting to explore
these connections in more detail and extend them to the case where higher derivative corrections are included in the effective action.

Finally, we note that the entropy current is conserved on-shell, but once we take the limit where the IR boundary coincides
with the horizon the variable $\phi^t$ decouples from the theory and the entropy current (which remains finite in this limit)
becomes conserved off-shell as well.

\section*{Acknowledgments}

We would like to thank Sayantani Bhattacharya, Stephen R. Green, Nima Arkani-Hamed, Veronika Hubeny,  Nabil Iqbal, Romuald Janik, Kristan Jensen, Ramalingam Loganayagam, Hong Liu, Mukund Rangamani, Dam T. Son, Andrei Starinets, Huan Yang, Andrzej Wereszczynski and Przemyslaw Witaszczyk for useful discussions, comments and correspondence. We would also like to thank the authors of \cite{Crossley:2015tka} for letting us know about their upcoming work. This work is part of the research programme of the Foundation for Fundamental Research on Matter (FOM), which is part of the Netherlands Organisation for Scientic Research (NWO). Research at Perimeter Institute is supported by the Government of Canada through Industry Canada and by the Province of Ontario through the Ministry of Research \& Innovation.

\bibliography{eff_biblio.bib}{}
\bibliographystyle{utphys}

\end{document}